\RequirePackage{setspace}
\documentclass[11pt]{article}%
\usepackage[linesnumbered,ruled,vlined]{algorithm2e}
\usepackage{algpseudocode} %algorithm

\usepackage{amssymb,amsfonts,amsmath,bm,mathrsfs}
\usepackage{dsfont}
\usepackage[left=1in,right=1in,top=1.25in,bottom=1.25in]{geometry}
\usepackage{multirow,booktabs,dcolumn,color}
\usepackage{graphicx,epstopdf}
\usepackage{import}
\usepackage[stable, multiple, splitrule, flushmargin]{footmisc}

\usepackage[para]{threeparttable}
\usepackage[sort&compress,longnamesfirst]{natbib}\bibpunct{(}{)}{;}{a}{}{,}

\usepackage{longtable, rotating, booktabs}%rotating has option figuresright so all turn the same way
\usepackage{eurosym}
\usepackage[font=small,format=plain,labelfont=bf,up,textfont=it,up]{caption}
\usepackage{subcaption}
\usepackage{multirow}
\usepackage{hyperref}
\hypersetup{
  colorlinks   = true, %Colours links instead of ugly boxes
  urlcolor     = blue, %Colour for external hyperlinks
  linkcolor    = blue, %Colour of internal links
  citecolor   = blue %Colour of citations
}

\usepackage{etoolbox}
\makeatletter
        \patchcmd{\NAT@citex}
          {\@citea\NAT@hyper@{%
             \NAT@nmfmt{\NAT@nm}%
             \hyper@natlinkbreak{\NAT@aysep\NAT@spacechar}{\@citeb\@extra@b@citeb}%
             \NAT@date}}
          {\@citea\NAT@nmfmt{\NAT@nm}%
           \NAT@aysep\NAT@spacechar\NAT@hyper@{\NAT@date}}{}{}

        \patchcmd{\NAT@citex}
          {\@citea\NAT@hyper@{%
             \NAT@nmfmt{\NAT@nm}%
             \hyper@natlinkbreak{\NAT@spacechar\NAT@@open\if*#1*\else#1\NAT@spacechar\fi}%
               {\@citeb\@extra@b@citeb}%
             \NAT@date}}
          {\@citea\NAT@nmfmt{\NAT@nm}%
           \NAT@spacechar\NAT@@open\if*#1*\else#1\NAT@spacechar\fi\NAT@hyper@{\NAT@date}}
          {}{}
\makeatother

\makeatletter %begins code block
\renewcommand{\paragraph}{\@startsection{paragraph}{4}{\z@}%\newcommand{cmd}[args][opt]{def}
{-2.5ex plus .2ex}%
{.01ex}%
{\normalfont\normalsize\itshape} }
\makeatother %ends code block

\newtheorem{theorem}{Theorem}

\newtheorem{lemma}{Lemma}

\mathchardef\mhyphen="2D
\makeatletter
\def\@biblabel#1{\hspace*{-\labelsep}}
\makeatother

\newcommand\fnsep{\textsuperscript{,}}
\newcommand\prob{\mathbb{P}}

\title{Surprised by the Hot Hand  Fallacy?\\ A Truth in the Law of Small Numbers
}

\author{Joshua B. Miller and Adam Sanjurjo
\thanks{\scriptsize Department of Economics (FAE), University of Alicante. Financial support from the Department of Decision Sciences at Bocconi University, and the Spanish Ministries of Education and Science and Economics and Competitiveness (ECO2015-65820-P) and Generalitat Valenciana (Research Projects Gruposo3/086 and PROMETEO/2013/037)
is gratefully acknowledged.}
\footnote{Both authors contributed equally, with names listed in alphabetical order.}
\footnote{This draft has benefitted from helpful comments and suggestions from the editor and anonymous reviewers, as well as Jason Abaluck, Jose Apesteguia, David Arathorn, Jeremy Arkes, Maya Bar-Hillel, Phil Birnbaum, Daniel Benjamin, Marco Bonetti, Colin Camerer, Juan Carrillo, Gary Charness, Ben Cohen, Vincent Crawford, Martin Dufwenberg, Jordan Ellenberg, Florian Ederer, Jonah Gabry, Andrew Gelman, Ben Gillen, Tom Gilovich, Maria Glymour, Uri Gneezy, Daniel Goldstein, Daniel Houser, Richard Jagacinski, Daniel Kahan, Daniel Kahneman, Erik Kimbrough, Dan Levin, Elliot Ludvig, Mark Machina, Daniel Martin, Filippo Massari, Guy Molyneux, Gidi Nave, Muriel Niederle, Christopher Olivola, Andreas Ortmann, Ryan Oprea, Carlos Oyarzun, Judea Pearl, David Rahman, Justin Rao, Alan Reifman, Pedro Rey-Biel, Yosef Rinott, Aldo Rustichini, Ricardo Serrano-Padial, Bill Sandholm, Vernon Smith, Lones Smith, Connan Snider, Joel Sobel, Charlie Sprenger, Daniel Stone, Sigrid Suetens, Dmitry Taubinsky, Richard Thaler, Michael J. Wiener, Nat Wilcox, and Bart Wilson. We would also like to thank seminar participants at Caltech, City U. London, Chapman U., Claremont Graduate School, Columbia U., Drexel U., George Mason U., New York University, NHH Norway, Max Planck Institute for Human Development (ABC), Microsoft Research, U. of Minnesota, Naval Postgraduate School, the Ohio State U., Santa Clara U., Stanford U., Tilburg U., U. de Alicante, U. del Pa\'{i}s Vasco, U. of Amsterdam,  UC Berkeley, UC Irvine, U. of Pittsburg, UC Santa Cruz, UC San Diego, U. New South Wales, U. Southern California, U. of Queensland, U. of Wellington, U. of Wisconsin, U. of Zurich, Washington State U., WZB Social Science Center, as well as conference participants at Gary's Conference, IMEBE Rome 2016,  M-BEES Maastricht 2015, SITE Stanford U. 2016, 11th World Congress of The Econometric Society, The 30th Annual Congress of the European Economic Association, and the 14th TIBER Symposium on Psychology and Economics. All mistakes and omissions remain our own.}
}
\date{October 17, 2018\\
\ \\
Published: {\it Econometrica}, Vol. 86, No. 6 (November 2018), 2019-2047 [\href{https://onlinelibrary.wiley.com/doi/abs/10.3982/ECTA14943}{link}]\\
}

\begin{document}
 %\beforepreface
  \maketitle
\begin{abstract}
\label{sec: Abstract}

We prove that a subtle but substantial bias exists in a common measure of the conditional dependence of present outcomes on streaks of past outcomes in sequential data. The magnitude of this {\it streak selection bias} generally decreases as the sequence gets longer, but increases in streak length, and remains substantial for a range of sequence lengths often used in empirical work. We observe that the canonical study in the influential hot hand fallacy literature, along with replications, are vulnerable to the bias. Upon correcting for the bias we find that the long-standing conclusions of the canonical study are reversed.
\end{abstract}



\clearpage

\setstretch{1.25}

%------------

\section{Introduction}\thispagestyle{plain}\label{sec: Introduction}

Jack the researcher takes a coin from his pocket and decides to flip it, say, one hundred times. As he is curious about what outcome typically follows a heads, whenever he flips a heads he commits to writing down the outcome of the next flip on the scrap of paper next to him. Upon completing the one hundred flips, Jack of course expects the proportion of heads written on the scrap of paper to be one-half. Shockingly, Jack is wrong. For a fair coin, the expected proportion of heads is smaller than one-half.

We prove that for any finite sequence of binary data, in which each outcome of ``success'' or ``failure'' is determined by an i.i.d. random variable, the proportion of successes among the outcomes that immediately follow a streak of consecutive successes is expected to be strictly less than the underlying (conditional) probability of success.\footnote{The expectation is conditional on the appearance of at least one streak of $k$ consecutive heads within the first $n-1$ trials, where $n\geq 3$ and $1\leq k<n-1$.} While the magnitude of this {\it streak selection bias} generally decreases as the sequence gets longer, it increases in streak length, and remains substantial for a range of sequence lengths often used in empirical work.

We observe that the canonical study in the influential {\it hot hand fallacy} literature,\footnote{The hot hand fallacy has been given considerable weight as a candidate explanation for various puzzles and behavioral anomalies identified in the domains of financial markets, sports wagering, casino gambling, and lotteries  \citep{RabinVayanos--RES--2010,Malkiel2011,DelongShleiferSummersWaldmann--JB--1991,LohWarachka--MS--2012,KahnemanRiepe--JPM--1998,deBondt--IJF--1993,BarberisThaler--HandbookoftheEconomicsofFinanceCh18--2003,Camerer--AER--1989,DurhamHertzelMartin--JF--2005,PaulWeinbach--JSE--2005,BrownSauer--AER--1993,SinkeyLogan--EEJ--2012,AveryChevalier--JB--1999,Arkes--wp--2011,LeeSmith--JBDM--2002,CrosonSundali--JRU--2005,SundaliCroson--JDM--2006,NarayananManchanda--QME--2012,SmithLevereKurtzman--MS--2009,XuHarvey--Cognition--2014,GuryanKearney--AER--2008,Galbo-JorgensenSuetensTyran--wp--2013,YuanSunSiu--JEBO--2014}. \label{fn: hot hand cites}
}  \citet{GilovichValloneTversky--CS--1985}, along with replications, have mistakenly employed a biased selection procedure that is analogous to Jack's.\footnote{For an extensive survey of the hot hand fallacy literature see \citep{MillerSanjurjo--HHSurvey--2017}.} Upon conducting a de-biased analysis we find that the long-standing conclusions of the canonical study are reversed.

To illustrate how the selection procedure that Jack uses in the opening example leads to a bias, consider the simplest case in which he decides to flip the coin three times, rather than 100. In this case there are only $2^3=8$ possibilities for the {\it single} three-flip sequence that Jack will observe. Column one of Table~\ref{tab: bias} lists these, with the respective flips that Jack would record (write down) underlined for each possible sequence. Column two gives the respective proportion of heads on recorded flips for each possible sequence. As Jack is equally likely to encounter each sequence, one can see that the expected proportion is strictly less than $1/2$, and in this case is 5/12.\footnote{The expectation is conditional on Jack having at least one flip to record.} Notice that because the sequence (rather than the flip) is the primitive outcome, the weight that the (conditional) expectation places on each sequence's associated proportion is independent of the number of recorded flips.\footnote{By contrast, if Jack were instead to observe \emph{multiple} sequences generated from the same coin, then he could weight each proportion according to its number of recorded flips when taking the average proportion across sequences. This would result in a relatively smaller bias that vanishes in the limit (see Appendix~\ref{ssec: Appendix consistency}).\label{fn: multiple sequences}}

\begin{table}[t]\centering
\caption{Column one lists the eight sequences that are possible for three flips of a fair coin. The proportion of heads on the flips that immediately follow one or more heads is reported in Column two, for each sequence that has at least one such flip. The (conditional) expectation of the proportion, which is simply its arithmetic average across the six equally likely sequences for which it is defined, is reported in the bottom row.\label{tab: bias}}
\scalebox{1}{
\begin{tabular}{ccc}
  \toprule
 Three flip &  Proportion of Hs \\
  sequence &  on recorded flips \\
  \midrule
  TTT &  -\\
   TTH &  -\\[3pt]
 TH\underline{T} & 0\\[3pt]
   H\underline{T}T & 0  \\
  TH\underline{H} & 1\\[3pt]
  H\underline{T}H & 0\\[3pt]
  H\underline{H}\underline{T} &  $\frac{1}{2}$  \\[3pt]
  H\underline{H}\underline{H} &  1  \\
  \midrule
 Expectation: &  $\frac{5}{12}$ \\
  \bottomrule
\end{tabular}
}
\end{table}

In Section~2 we prove the existence of the streak selection bias for the general case, then quantify it with a formula that we provide. In the case of streaks of length $k=1$ (as in the examples discussed above) the formula admits a simple representation, and the bias is tightly related to a form of finite sample bias that shows up in autoregressive coefficient estimators \citep{Yule--JRSS--1926,ShamanStine--JASA--1988}.\footnote{In the context of time series regression this bias is known as the {\it Hurwicz bias} \citep{Hurwicz1950}, which is exacerbated when one introduces fixed effects into a time series model with few time periods \citep{Nerlove1967,Nerlove--Econometrica--1971,Nickell--Econometrica--1981}. In Web Appendix~\ref{ssec: Appendix SWOR, Berkson, FSB}, we use a sampling-without-replacement argument to show that in the case of $k=1$ the streak selection bias, along with finite sample bias for autocorrelation (and time series), are essentially equivalent to: (i) a form of selection bias known in the statistics literature as Berkson's bias, or Berkson's paradox \citep{Berkson--BB--1946,RobertsSpitzerDelmoreSackett--JCD--1978}, and (ii) several classic conditional probability puzzles.} By contrast, for the more general case of $k>1$ the streak selection bias is typically of larger magnitude, and the formula does not appear to admit a simple representation.\footnote{In Web Appendix~\ref{sec: Web Appendix bias Bias Mechanism} we show that the bias can be decomposed into two factors: a form of sampling-without-replacement, and a stronger bias driven by the overlapping nature of the selection procedure. In Web Appendix~\ref{ssec: Web Appendix overlap k>1} we show how the bias due to the overlapping nature of the selection procedure is related to the {\it overlapping words paradox} \citep{GuibasOdlyzko--JCTA--1981}.} In this case we provide a formula for the bias that is numerically tractable for sequence lengths commonly used in the literature that we discuss.

The bias has important implications for the analysis of streak effects in the hot hand fallacy literature. The fallacy refers to the conclusion of the seminal work of Gilovich, Vallone, and Tversky~(1985; henceforth GVT), in which the authors found that despite the near ubiquitous belief among basketball fans and experts that there is momentum in shooting performance (``hot hand'' or ``streak'' shooting) the conclusion from their statistical analyses was that momentum did not exist.\footnote{In particular, they observed that basketball shooting is ``analogous to coin tossing'' and ``adequately described by a simple binomial model.'' From this, they concluded that the belief in the hot hand was both ``erroneous'' and ``a powerful and widely shared cognitive illusion'' \citep[pp.312--313]{GilovichValloneTversky--CS--1985}}  The result has long been considered a surprising and stark exhibit of irrational behavior, as professional players and coaches have consistently rejected the conclusion, and its implications for their decision making. Indeed, in the years since the seminal paper was published a consensus has emerged that the hot hand is a ``myth,'' and the associated belief a ``massive and widespread cognitive illusion'' \citep{ThalerSunstein2008,Kahneman2011}.

We find that GVT's critical test of hot hand shooting is vulnerable to the bias for the following simple reason: just as it is (surprisingly) incorrect to expect a fair coin flipped 100 times to yield heads half of the time on those flips that immediately follow three consecutive heads, it is incorrect to expect a consistent 50 percent (Bernoulli i.i.d.) shooter who has taken 100 shots to make half of the shots that immediately follow a streak of three hits. Thus, after first replicating the original results using GVT's: (i) raw data, (ii) biased measures, and (iii) statistical tests, we perform a bias correction to GVT's measures, then repeat their statistical tests. We also run some additional (unbiased) tests as robustness checks. In contrast with GVT's results, the bias-corrected re-analysis reveals significant evidence of streak shooting, with large effect sizes.

In a brief discussion of the related literature in Section \ref{sec: HF}, we first observe that the two replications of GVT \citep{AvugosBar-EliRitovSher--IJSEP--2013,KoehlerConley--JSEP--2003} are similarly vulnerable to the bias. We illustrate how the results of \cite{AvugosBar-EliRitovSher--IJSEP--2013}, a close replication of GVT, similarly reverse when the bias is corrected for. \cite{MillerSanjurjo2015d} show that the results of \cite{KoehlerConley--JSEP--2003}, which has been referred to as ``an ideal situation in which to study the hot hand'' \citep{ThalerSunstein2008}, reverse when an unbiased (and more powered) analysis is performed. These results in turn agree with the unbiased analyses performed on all remaining extant controlled shooting datasets in \cite{MillerSanjurjo--wp--2014}. Conservative estimates of hot hand effect sizes are consistently moderate to large across studies.

It follows from these results that the hot hand is not a myth, and that the associated belief is not a cognitive illusion. In addition, because researchers have: (i) accepted the null hypothesis that players have a fixed probability of success, and (ii) treated the {\it mere} belief in the hot hand as a cognitive illusion, the hot hand fallacy itself can be viewed as a fallacy.\footnote{While our evidence reveals that belief in the hot hand is not a fallacy, it remains possible that those who believe in the hot hand hold beliefs that are too strong (or too weak), or cannot accurately detect the hot hand when it occurs. In Section~\ref{ssec: Belief in the hot hand} we briefly discuss existing evidence on beliefs.}

Finally, because the bias is subtle and (initially) surprising,  even for people well-versed in probability and statistics, those unaware of it may be susceptible to being misled, or exploited.\footnote{In informal conversations with researchers, and surveys of students, we have found a near-universal belief that the sample proportion should be equal to the underlying probability, in expectation.  The conviction with which these beliefs are often held is notable, and reminiscent of the arguments that surrounded the classic Monty Hall problem \citep{Friedman1998,Selvin--AS--1975,Nalebuff--JEP1--1987,VosSavant--ParadeMagazine--1990}. See \cite{MillerSanjurjo2015c} for more details on the connection between the selection bias, the Monty Hall problem, and other conditional probability puzzles.}
On the most basic level, it is possible that a na{\"i}ve observer could be convinced that negative sequential dependence exists in an i.i.d. random process if sample size information (i.e. the number of flips that Jack records) is obscured.\footnote{In particular, \citet{MillerSanjurjo--WorkingPaper--2016} show that the bias introduced here, in conjunction with a quasi-Bayesian model of decision making under sample size neglect \citep{GriffinTversky--CP--1992,KahnemanTversky--CP--1972,BenjaminRabinRaymond--WP--2014}, provides a novel structural candidate explanation for the persistence of gambler's fallacy beliefs.} More subtly, the bias can be leveraged to manipulate people into believing that the outcomes of an unpredictable process can be predicted at rates better than chance.\footnote{For example, suppose that a predictor observes successive realizations from a binary (or binarized) i.i.d. random process (e.g. daily stock price movements), and is evaluated according to the {\it success rate} of her predictions over, say, three months. If the predictor is given the freedom of {\it when} to predict, then she can exceed chance in her expected success rate simply by predicting a reversal whenever there is a streak of consecutive outcomes of the same kind.} Lastly, the bias can be applied in a straightforward way to construct gambling games that appear actuarially fair, but are not.\footnote{A simple example is to sell the following lottery ticket for \$5. A fair coin will be flipped 4 times. For each flip the outcome will be recorded if and only if the previous flip is a heads. If the proportion of recorded heads is strictly greater than one-half then the ticket pays \$10; if the proportion is strictly less than one-half then the ticket pays \$0; if the proportion is exactly equal to one-half, or if no flip is immediately preceded by a heads, then a new sequence of 4 flips is generated. While, intuitively, it seems that the expected value of the lottery must be \$5, it is instead \$4.}

\section{The Streak Selection Bias}\label{sec: Bias Section}

Let $\bm{X}=\{X_i\}_{i=1}^n$ be a sequence of binary random variables, with $X_i=1$  a ``success'' and $X_i=0$ a ``failure.''  A natural procedure for estimating the probability of success on trial $t$, conditional on trial $t$ immediately following $k$ consecutive successes, is to first select the subset of trials that immediately follow $k$ consecutive successes  $I_{k}(\bm{X}):=\{ i: {\textstyle\prod_{j=i-k}^{i-1}} X_j =1\}\subseteq \{k+1,\dots,n\}$, then calculate the proportion of successes on these trials.\footnote{In fact, this procedure yields the maximum likelihood estimate for $\prob(X_t=1\ | \ \prod_{j=t-k}^{t-1} X_j=1)$.} The following theorem establishes that when $\{X_i\}_{i=1}^n$ is a sequence of i.i.d random variables, with probability of success $p$ and fixed length $n$, this procedure yields a biased estimator of the conditional probability, $\prob(X_t=1\ | \ \prod_{j=t-k}^{t-1} X_j=1)\equiv p$.

\begin{theorem}\label{thm: General bias}
Let $\bm{X}=\{X_i\}_{i=1}^n$, $n\geq 3$, be a sequence of independent Bernoulli trials, each with probability of success $0<p<1$. Let $\hat{P}_{k}(\bm{X})$ be the proportion of successes on the subset of trials $I_{k}(\bm{X})$ that immediately follow $k$ consecutive successes, i.e. $\hat{P}_{k}(\bm{X}):=\sum_{i\in I_{k}(\bm{X})} X_i/|I_{k}(\bm{X})|$. $\hat{P}_{k}$ is a biased estimator of $\prob(X_t=1\ | \ \prod_{j=t-k}^{t-1} X_j=1)\equiv p$ for all $k$ such that $1\leq k\leq n-2$. In particular,

\begin{equation}
E\left.\left[\ \hat{P}_{k}(\bm{X}) \ \right| \ I_{k}(\bm{X}) \neq \emptyset     \ \right]< p
\end{equation}
\end{theorem}
\textbf{Outline of Proof}: In the proof contained in Appendix~\ref{sec: Appendix general Proofs} we begin by showing that the conditional expectation $E[  \hat{P}_{k}(\bm{X}) | I_{k}( \bm{X}) \neq \emptyset   ]$ is equal to the conditional probability $\prob(X_\tau=1| I_{k}(\bm{X}) \neq \emptyset)$, where $\tau$ is a trial drawn (uniformly) at random from the set of selected trials $I_{k}(\bm{X})$. Next, we show that for all eligible trials $t\in I_{k}(\bm{X})$ we have that $\prob(X_t=1| \tau=t, I_{k}(\bm{X}) \neq \emptyset)\leq p$,  with the inequality strict for $t<n$, which implies that  $\prob(X_\tau=1| I_{k}(\bm{X}) \neq \emptyset)<p$.  The strict inequality for $t<n$ follows from an application of Bayes' rule. In particular, we observe that $\prob(X_t=1| \tau=t, I_{k}(\bm{X}) \neq \emptyset)= \prob(X_t=1| \tau=t, \prod_{i=t-k}^{t-1}X_i=1)\propto \prob(\tau=t|X_t=1,\prod_{i=t-k}^{t-1}X_i=1)\times\prob(X_t=1|\prod_{i=t-k}^{t-1}X_t=1)=\prob(\tau=t|X_t=1,\prod_{i=t-k}^{t-1}X_t=1)\times p$, and then argue that $\prob(\tau=t|X_t=1,\prod_{i=t-k}^{t-1}X_i=1)< \prob(\tau=t|X_t=0,\prod_{i=t-k}^{t-1}X_i=1)$ for $t<n$, which guarantees that the likelihood ratio (updating factor) is less than one, and yields  $\prob(X_t=1| \tau=t, \prod_{i=t-k}^{t-1}X_i=1)<p$ for $t<n$.
The intuition for why $\tau =t$ is more likely when $X_t=0$ is the following: because the streak of ones ($\prod_{i=t-k}^{t-1}X_i=1$) is interrupted by $X_t=0$, the next $k$ trials are necessarily excluded from the set $I_{k}(\bm{X})$. This means that when $X_t=0$ there are, on average, fewer eligible trials in $I_{k}(\bm{X})$ from which to draw  (relative to when $X_t=1$), which implies that any single trial is more likely to be drawn.

In Web Appendix~\ref{sec: Web Appendix bias Bias Mechanism} we show that the downward bias can be decomposed into two factors: (i) {\it sampling-without-replacement}: the restriction that the finite number of available successes places on the procedure for selecting trials into $I_k(\bm{X})$, and (ii) {\it streak overlap}: the additional, and stronger, restriction that the arrangement of successes and failures in the sequence places on the procedure for selecting trials into $I_k(\bm{X})$.

Though $\hat{P}_{k}(\bm{X})$ is biased, it is straightforward to show that it is a consistent estimator of $\prob(X_t=1\ | \ \prod_{j=t-k}^{t-1} X_j=1)$.\footnote{See Appendix~\ref{ssec: Appendix consistency} for a proof.}\fnsep\footnote{It is possible to devise alternative estimators of the conditional probability that are unbiased. To illustrate, if the researcher were instead to control the number of selected trials by repeating the experiment until he generates exactly $m$ trials that immediately follow $k$ consecutive successes, then the proportion would be unbiased. Alternatively, if the researcher were to eliminate the overlapping nature of the measure, there would be no bias, even though the number of selected trials would still be random. In particular, for a sequence of $n$ trials, one can take each run of successes, and if it is of even length $2\ell$, divide it into blocks of two trials; if it is of odd length $2\ell -1$ include the right adjacent tails and divide it into blocks of two trials. In each case, the run of successes contributes $\ell$ observations. \label{fn: inverse sampling and overlap}}

\subsection{Quantifying the bias.}\label{ssec: Quantifying the bias}

\begin{figure}[t]
\begin{centering}
  \includegraphics[height=4.5in]{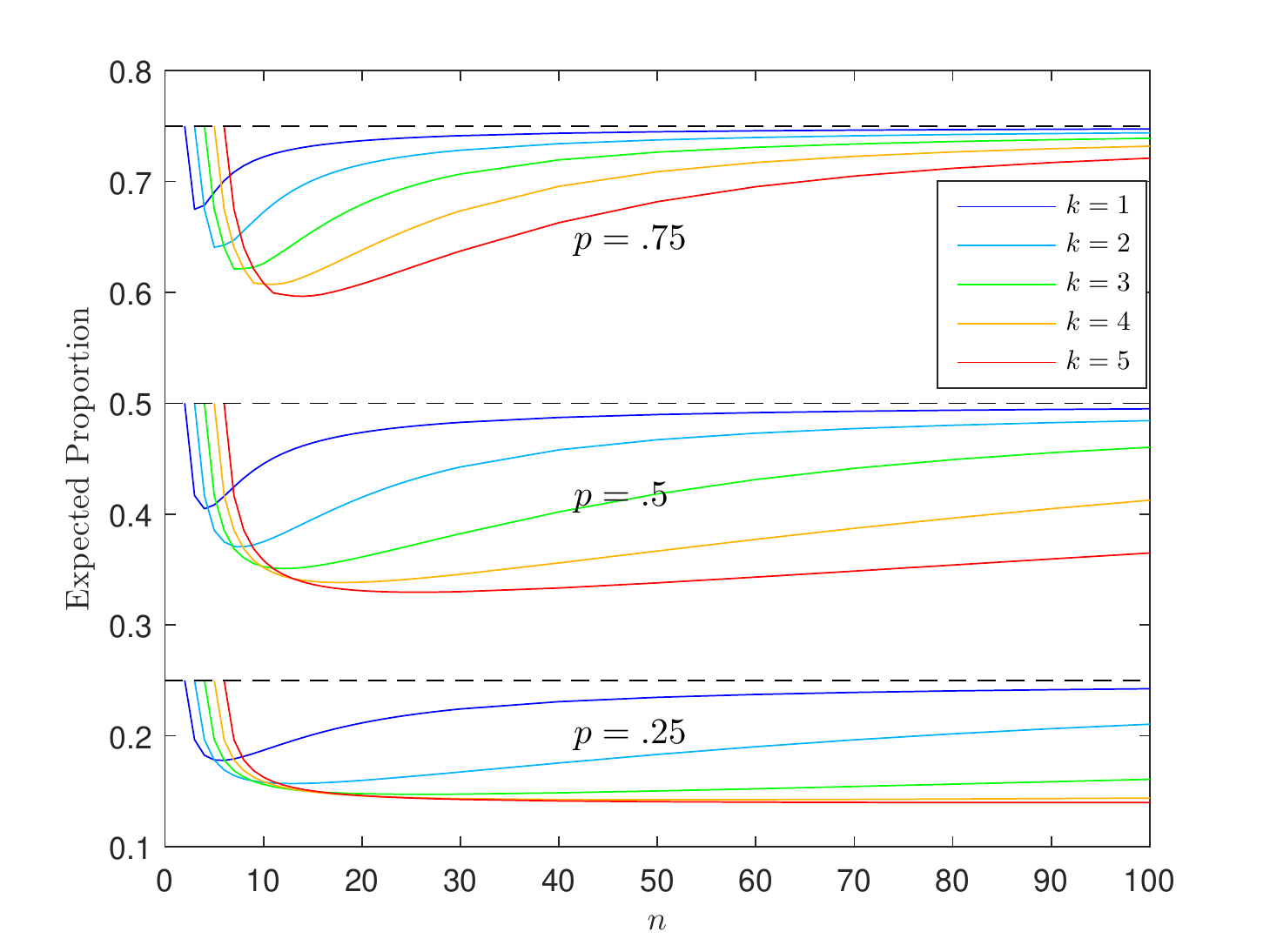}
  \caption{The expected value of the proportion of successes on trials that immediately follow $k$ consecutive successes, $\hat{P}_{k}(\bm{X})$, as a function of the total number of trials $n$, for different values of $k$ and probabilities of success $p$, using the formula provided in Web Appendix~\ref{sec: Appendix recursive}.}
  \label{fig: Bias Prop}
\end{centering}
\end{figure}

In Web Appendix~\ref{ssec: Appendix recursive diff} we provide a formula that can be used to calculate $E[\ \hat{P}_{k}(\bm{X}) \ | \  I_{k}(\bm{X}) \neq \emptyset     \ ]$. For the special case of $k=1$ a closed form exists, which we provide in Appendix~\ref{sec: Appendix k=1 Proofs}.  There does not appear to be a simple representation for $k>1$.

Figure~\ref{fig: Bias Prop} contains a plot of $E[\ \hat{P}_{k}(\bm{X}) \ | \  I_{k}(\bm{X}) \neq \emptyset     \ ]$, as a function of the number of trials in the sequence $n$, and for different values of $k$ and $p$.  The dotted lines in the figure represent the true probability of success for $p=0.25, 0.50$, and $0.75$, respectively. The five solid lines immediately below each dotted line represent the respective expected proportions for each value of $k=1,2,\dots,5$. Observe that while the bias does generally decrease as $n$ increases, it can remain substantial even for long sequences. For example, in the case of $n=100$, $p=0.5$, and $k=5$, the magnitude of the bias is $.35 - .50= -0.15$, and in the case of $n=100$, $p=0.25$, and $k=3$, the magnitude of the bias is $.16-.25 = - 0.09$.\footnote{The non-monotonicity in $n$ of the curves presented in Figure~\ref{fig: Bias Prop} arises because for any streak length $k$ there is no bias when $n=k+1$ (because there are only two feasible sequences, which are equally likely), or in the limit (see Appendix~\ref{ssec: Appendix consistency}).}

\section{Application to the Hot Hand Fallacy}\label{sec: HF}

\begin{quote}
{\it This account explains both the formation and maintenance of the erroneous belief in the hot hand: if random sequences are perceived as streak shooting, then no amount of exposure to such sequences will convince the player, the coach, or the fan that the sequences are in fact random.} (Gilovich, Vallone, and Tversky [GVT] 1985)
\end{quote}

In their seminal paper GVT find no evidence of hot hand shooting in their analysis of basketball shooting data, despite the near-unanimous belief in the hot hand among players, coaches, and fans. As a result, they conclude that belief in the hot hand is a ``powerful and widely shared cognitive illusion.'' (p. 313).

\subsection{GVT's analysis}

\paragraph{Empirical approach}
GVT's ``Analysis of Conditional Probabilities'' is their main test of hot hand shooting, and provides their only measure of the magnitude of the hot hand effect.  The goal of their analysis is to determine whether a player's hit probability is higher following a streak of hits than it is following a streak of misses.\footnote{GVT explicitly treat hot hand and streak shooting as synonymous \citep[pp. 296-297]{GilovichValloneTversky--CS--1985}. \citet{MillerSanjurjo--wp--2014} provide an analysis that distinguishes between hot hand and cold hand shooting, and find hot hand shooting across all extant controlled shooting datasets, but little in the way of cold hand shooting. Thus, in the present analysis we use the terms streakiness and hot hand shooting interchangeably.} To this end, GVT reported each player $i$'s shooting  percentage conditional on having: (1) hit the last $k$ shots, $\hat{P}^i(hit| k\ hits)$, and  (2) missed the last $k$ shots, $\hat{P}^i(hit| k\ misses)$, for streak lengths $k=1,2,3$ (Table 4, p. 307).\footnote{We abuse our notation from Section \ref{sec: Bias Section} here in order to facilitate comparison with GVT's analysis: we use $\hat{P}^i(hit| k\ hits)$ for both the random variable $\hat{P}_k(\bm{X})$ and its realization $\hat{P}_k(\bm{x})$. Similarly, we use $\hat{P}^i(hit| k\ misses)$ for the proportion of successes on trials that immediately follow $k$ consecutive failures. }
After informally comparing these shooting percentages for individual players, GVT performed a paired t-test of whether $E[\hat{P}^i(hit| k\ hits)-\hat{P}^i(hit| k\ misses)]=0$, for $k=1,2,3$.\footnote{Under the null hypothesis the difference between each $i$'s pair of shooting percentages is drawn from a normal distribution with mean zero.}\fnsep\footnote{While GVT's analysis of conditional probabilities provides their only measure of the magnitude of the hot hand, they also analyze the number of runs, serial correlation, and variation of shooting percentage in 4-shot windows.  \citet{MillerSanjurjo--wp--2014} show that the runs and serial correlation tests, along with the conditional probability test for $k=1$, all amount to roughly the same test, and moreover, that they are not sufficiently powered to identify hot hand shooting. The reason why is due to measurement error: the act of hitting a single shot is only a weak signal of a change in a player's underlying probability of success, which leads to an attenuation bias in the estimate of the increase in the  probability of success associated with entering the hot state (see Appendix~\ref{sec: Appendix Bias adjustment} and \citet{Stone--AS--2012}'s work on measurement error when estimating autocorrelation in ability). The test of variation in 4-shot windows is even less powered than the aforementioned tests \citep{Wardrop--wp--1999,MillerSanjurjo--wp--2014}.\label{fn: GVTs 3 serial corr}}

In the remainder of this section, we focus our discussion on streaks of length three (or more), as in, e.g. \cite{KoehlerConley--JSEP--2003,Rao--wp--2009b}, given that: (i) shorter streak lengths exacerbate attenuation bias due to measurement error (see Footnote~\ref{fn: GVTs 3 serial corr} and Appendix~\ref{sec: Appendix Bias adjustment}), and (ii) people typically perceive streaks as beginning with the third successive event \citep{CarlsonShu--OBHDP--2007}.  In any case, robustness checks using different streak lengths yield similar results (see Footnotes \ref{fn: robust to varying k.} and \ref{fn: robust to varying k permutation test} in Section \ref{ssec: debiasing GVTs test}).

\paragraph{Data}

GVT analyze shot sequences from basketball players in three contexts: NBA field goal data, NBA free-throw data, and a controlled shooting experiment with NCAA collegiate players. The shooting experiment was GVT's controlled test of hot hand shooting, designed for the purpose of ``eliminating the effects of shot selection and defensive pressure'' (p. 34), which makes it central to their main conclusions. Thus, we focus on this data below when discussing the relevance of the bias to GVT's results.\footnote{From the statistical point of view, the in-game field goal data that GVT analyze (Study 2: 76ers, 1980-81 season: 9 players, 48 home games) is not ideal for the study of hot hand shooting for reasons that are unrelated to the bias. The most notable concern with in-game field goal data is that the opposing team has incentive to make \emph{costly} strategic adjustments to mitigate the impact of the ``hot'' player \citep[p. 17]{DixitNalebuff1991}. This concern has been emphasized by researchers in the hot hand literature \citep{AharoniSarig--AE--2011,GreenZwiebel--wp--2013}, and is not merely theoretical, as it has a strong empirical basis.  While GVT observed that a shooter's field goal percentage is lower after consecutive successes, subsequent studies have shown that with even partial controls for defensive pressure (and shot location), this effect is eliminated \citep{Rao--wp--2009a,BocskocskyEzekowitzStein--MIT--2014}. Further, evidence of specific forms of strategic adjustment has been documented \citep{AharoniSarig--AE--2011,BocskocskyEzekowitzStein--MIT--2014}. See \citet{MillerSanjurjo--wp--2014} for further details.}\fnsep\footnote{The in-game free throw data that GVT analyze (Study 3: Celtics, 1980-81, 1981-82 seasons: 9 players), while arguably controlled, is not ideal for the study of hot hand shooting for a number of reasons: (i) hitting the first shot in a pair of isolated shots is not typically regarded by fans and players as hot hand shooting \citep{KoehlerConley--JSEP--2003}, presumably due to the high prior probability of success ($\approx .75$),  (ii) hitting a single shot is a weak signal of a player's underlying state, which can lead to severe measurement error \citep{Stone--AS--2012,Arkes--JSE--2013}, (iii) it is vulnerable to an omitted variable bias, as free throw pairs are relatively rare, and shots must be aggregated across games and seasons in order to have sufficient sample size \citep{MillerSanjurjo--wp--2014}. In any event, subsequent studies of free throw data have found evidence that is inconsistent with the conclusions that GVT drew from the Celtics' data \citep{Wardrop--AS--1995,Arkes--JQAS--2010,YaariEisenmann--PLOSONE--2011,AharoniSarig--AE--2011, GoldmanRao--MIT--2012,MillerSanjurjo--wp--2014}.}

In GVT's controlled shooting experiment 26 players from the Cornell University Mens' (14) and Womens' (12) basketball teams participated in an incentivized shooting task. Each player shot 100 times at a distance from which the experimenters determined he/she would make around 50 percent of the shots. Following each shot the player had to change positions along two symmetric arcs---one facing the basket from the left, and the other from the right.

\paragraph{Results}
In Columns 4 and 5 of Table \ref{tab: CpDiffCorrected3+GVTv3} we use the raw data from GVT to reproduce the shooting percentages, $\hat{P}^i(hit| 3\ hits)$ and $\hat{P}^i(hit| 3\ misses)$, for each of the 26 players (these are identical to Columns 2 and 8 of Table~4 in GVT). As indicated in GVT, players on average hit .49 when on a hit streak, versus .45 when on a miss streak. GVT's paired t-test finds the difference to be statistically indistinguishable from zero, and we replicate this result ($p=.49$).

\subsection{The bias in GVT's analysis}\label{ssec: HH bias in GVT}

\begin{table}[htbp]\centering
\def\sym#1{\ifmmode^{#1}\else\(^{#1}\)\fi}
\caption{Columns~4 and 5 reproduce the shooting percentages and number of shots that appear in Table~4, Columns~2 and 8, from \citet{GilovichValloneTversky--CS--1985} (note: 3 hits (misses) includes streaks of 3, 4, 5, etc.). Column 6 reports the difference between the proportions (using the raw data), and column 7 adjusts for the bias (mean correction), based on each player's shooting percentage (probability in this case) and number of shots.  \label{tab: CpDiffCorrected3+GVTv3}}
\begin{tabular}{r*{1}{
c
D{.}{.}{0.3}
c
D{.}{.}{0.3}D{.}{.}{0.3}
c
D{.}{.}{0.3}
D{.}{.}{0.3}
c
}}
\toprule
         & & &\multicolumn{2}{c}{}& & &\multicolumn{2}{c}{\footnotesize  $\hat{D}_3:=\hat{P}(hit|3 \ hits)-\hat{P}(hit| 3\ misses)$} &                                   \\
                     \cmidrule{8-9}
Player             &\multicolumn{1}{c}{\# shots} &\multicolumn{1}{c}{$\hat{P}(hit)$}&            &\multicolumn{1}{c}{\footnotesize $\hat{P}(hit|3 \ hits)$}&\multicolumn{1}{c}{\footnotesize $\hat{P}(hit| 3\ misses)$}&            &\multicolumn{1}{c}{GVT est.}&\multicolumn{1}{c}{bias adj.}            \\
\midrule
Males &&&&&&&&\\
1                 &          100 &         .54&            &         .50\ (12)&         .44\ (9)&            &         .06&         .14&            \\
2                 &          100 &         .35&            &         .00\ (3)&         .43\ (28)&            &        -.43&        -.33&            \\
3                 &          100 &         .60&            &         .60\ (25)&         .67\ (6)&            &        -.07&         .02&            \\
4                 &          90  &         .40&            &         .33\ (3)&         .47\ (15)&            &        -.13&        -.03&            \\
5                 &          100 &         .42&            &         .33\ (6)&         .75\ (12)&            &        -.42&        -.33&            \\
6                 &          100 &         .57&            &         .65\ (23)&         .25\ (12)&            &         .40&         .48&            \\
7                 &          75  &         .56&            &         .65\ (17)&         .29\ (7)&            &         .36&         .47&            \\
8                 &          50  &         .50&            &         .57\ (7)&         .50\ (6)&            &         .07&         .24&            \\
9                 &          100 &         .54&            &         .83\ (30)&         .35\ (20)&            &         .48&         .56&             \\
10                 &         100 &         .60&            &         .57\ (21)&         .57\ (7)&            &         .00&         .09&           \\
11                 &         100 &         .58&            &         .62\ (21)&         .57\ (7)&            &         .05&         .14&           \\
12                 &         100 &         .44&            &         .43\ (7)&         .41\ (17)&            &         .02&         .10&           \\
13                 &         100 &         .61&            &         .50\ (18)&         .40\ (5)&            &         .10&         .19&           \\
14                 &         100 &         .59&            &         .60\ (20)&         .50\ (6)&            &         .10&         .19&           \\
Females &&&&&&&&\\
1                 &          100 &         .48&            &         .33\ (9)&         .67\ (9)&            &        -.33&        -.25&            \\
2                 &          100 &         .34&            &         .40\ (5)&         .43\ (28)&            &        -.03&         .07&            \\
3                 &          100 &         .39&            &         .50\ (8)&         .36\ (25)&            &         .14&         .23&            \\
4                 &          100 &         .32&            &         .33\ (3)&         .27\ (30)&            &         .07&         .17&            \\
5                 &          100 &         .36&            &         .20\ (5)&         .22\ (27)&            &        -.02&         .08&            \\
6                 &          100 &         .46&            &         .29\ (7)&         .54\ (11)&            &        -.26&        -.18&            \\
7                 &          100 &         .41&            &         .62\ (13)&         .32\ (25)&            &         .30&         .39&            \\
8                 &          100 &         .53&            &         .73\ (15)&         .67\ (9)&            &         .07&         .15&            \\
9                 &          100 &         .45&            &         .50\ (8)&         .46\ (13)&            &         .04&         .12&            \\
10                 &         100 &         .46&            &         .71\ (14)&         .32\ (19) &            &         .40&         .48&           \\
11                 &         100 &         .53&            &         .39\ (13)&         .50\ (10) &            &        -.12&        -.04&           \\
12                 &         100 &         .25&            &           .-\ \ (0)&        .32\ (37)&            &           .&           .&           \\
\midrule
Average               &             &         .47&            &         .49&         .45&            &         .03&         .13&         \\
\bottomrule
\end{tabular}
\end{table}

While GVT's null hypothesis that $E[\hat{P}^i(hit| k\ hits)-\hat{P}^i(hit| k\ misses)]=0$ seems intuitively correct for a consistent shooter with a fixed probability of success $p^i$ (i.i.d. Bernoulli), Theorem~\ref{thm: General bias} reveals a flaw in this reasoning.  In particular, we have established that $\hat{P}^i(hit| k\ hits)$ is expected to be less than $p^i$, and $\hat{P}^i(hit| k\ misses)$ greater than $p^i$ (by symmetry). In fact, in Appendix~\ref{ssec: Appendix exp diff in prop} we show that the difference $\hat{P}^i(hit| k\ hits)-\hat{P}^i(hit| k\ misses)$ is not only expected to be negative, but that its magnitude is more than double the bias in either of the respective proportions.\footnote{That the difference is expected to be negative does not follow immediately from Theorem~\ref{thm: General bias}, as the set of sequences for which the difference is well-defined is a strict subset of the set corresponding to either of the respective proportions.  Nevertheless, the reasoning of the proof is similar. See Theorem~\ref{thm: General bias diff} of Appendix~\ref{ssec: Appendix exp diff in prop}.}

Under GVT's design target of each player taking $n=100$ shots and making half ($p=.5$) of them, we use the results from Section \ref{sec: Bias Section} and Appendix \ref{ssec: Appendix exp diff in prop} to find that the expected difference (and the strength of the bias) is -8 percentage points.\footnote{See Figure \ref{fig: BiasDiffProp} in Appendix \ref{ssec: Appendix exp diff in prop} for the bias in the difference as $n,p$ and $k$ vary.} Therefore, the difference between the average proportion of +4 percentage points observed by GVT is actually +12 percentage points higher than the difference that would be expected from a Bernoulli i.i.d. shooter. Thus, the bias has long disguised evidence in GVT's data that may well indicate hot hand shooting.

\subsection{A bias-corrected statistical analysis of GVT}\label{ssec: debiasing GVTs test}

A straightforward way to adjust for the bias in GVT's analysis is simply to shift the difference for each shooter by the amount of the corresponding bias, then repeat their paired t-test. While this test yields a statistically significant result ($p<.05$), the paired t-test limits statistical power because it reduces each player's performance to a single number, ignoring the number of shots that the player attempted in each category, i.e. ``3 hits'' and ``3 misses.''  In addition, adjusting for the bias based on the assumption that $p=.5$ assumes that GVT's design target was met precisely.

As a result, for each player we again compute the bias under the null hypothesis that trials are i.i.d. Bernoulli (i.e. ``consistent'' shooting) but now with a probability of success equal to the player's observed shooting percentage (Column 3 of Table~\ref{tab: CpDiffCorrected3+GVTv3}), and using the number of shots taken in each category to inform our standard errors. With this approach the average difference goes from +3 to a considerable +13 percentage points ($p<.01$, $S.E.=4.7$pp).\footnote{The standard error is computed based on the assumption of independence across the 2600 trials, and normality. In particular, defining player $i$'s difference $\hat{D}^i_k:=\hat{P}^i(hit| k\ hits)-\hat{P}^i(hit| k\ misses)$, the variance satisfies $\widehat{Var}(\hat{D}^i_k)=\widehat{Var}(\hat{P}^i(hit| k\ hits))+\widehat{Var}(\hat{P}^i(hit| k\ misses))$ for each player $i$.  Simulations reveal that the associated $(1-\alpha)\times 100\%$ confidence intervals with radius $z_{\alpha/2}\times\widehat{Var}(\bar{D}_k)^{1/2}$ (where the mean difference is given by $\bar{D}_k:=(1/n)\sum_{i=1}^n \hat{D}^i_k$) have the appropriate coverage---i.e. $(1-\alpha/2)\times 100\%$ of the time the true difference is greater than $\bar{D}_k-z_{\alpha/2}\times\widehat{Var}(\bar{D}_k)^{1/2}$, for both Bernoulli trials and the positive feedback model discussed in Section~\ref{sec: Appendix Bias adjustment}.\label{fn: st.errors for normal approx}}\fnsep\footnote{For an alternative approach that involves pooling shots across players, and yields similar results, see Appendix \ref{sec: Appendix more from sec 3}.} To put the magnitude of +13 percentage points into perspective, the difference between the median three point shooter and the top three point shooter in the 2015-2016 NBA season was 12 percentage points.\footnote{ESPN, ``NBA Player 3-Point Shooting Statistics - 2015-16.''   \href{http://www.espn.com/nba/statistics/player/_/stat/3-points}{ http://www.espn.com/nba/statistics/player/\_/stat/3-points} [accessed September 24, 2016].} Further, this is a {\it conservative} estimate because in practice the data generating processes (i.e. shooters) clearly differ from i.i.d. Bernoulli trials, and the bias becomes much larger under various models of hot hand shooting because of measurement error (see Appendix~\ref{sec: Appendix Bias adjustment}).

\begin{figure}[t]
\begin{centering}
  \includegraphics[height=3.5in]{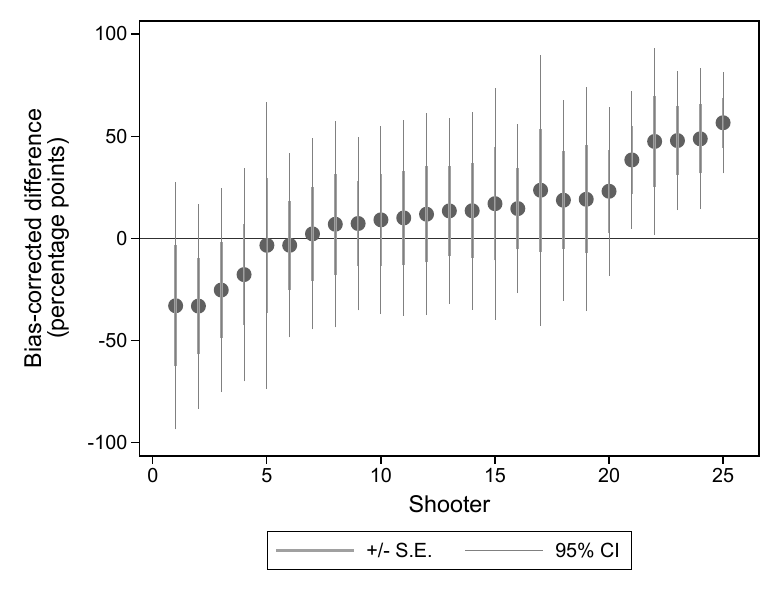}
  \caption{The bias-corrected difference  $\hat{D}^i_3=\hat{P}^i(hit| 3\ hits)-\hat{P}^i(hit| 3\ misses)$ for each player, under the assumption that his/her probability of success is equal to his/her overall shooting percentage.}
  \label{fig: Diff25players}
\end{centering}
\end{figure}

GVT also informally discussed the heterogeneity across players, and asserted that most players shot relatively better when on a streak of misses than when on a streak of hits. By contrast, Figure~\ref{fig: Diff25players} shows that once the bias correction is made to the differences 19 of the 25 players directionally exhibit hot hand shooting, which is itself significant ($p<.01$, binomial test).\footnote{Repeating the tests for longer ($k=4$) or shorter ($k=2$) streak lengths yields similar results that are consistent with the attenuation bias in estimated effect sizes discussed in Footnote~\ref{fn: GVTs 3 serial corr}. In particular, If we instead define a streak as beginning with 4 consecutive hits, which is a stronger signal of hot hand shooting, then the average bias-adjusted difference in proportions is 10 percentage points ($p=.07$, $S.E.=6.9$, one-sided test), and four players exhibit statistically significant hot hand shooting ($p<.05$), which is itself significant ($p<.01$, binomial test). On the other hand, if we define a streak as beginning with 2 consecutive hits, which is a weaker signal of hot hand shooting, then the average bias-adjusted difference in proportions is $5.4$ percentage points ($p<.05$, $S.E.=3$, one-sided test), and four players exhibit statistically significant hot hand shooting ($p<.05$), which is itself significant ($p<.01$, binomial test).\label{fn: robust to varying k.}} Further, as indicated by the confidence intervals, t-tests reveal that 5 of the players exhibit statistically significant evidence of hot hand shooting ($p<.05$, t-test), which, for a set of 25 independent tests, is itself significant ($p<.01$, binomial test).

\paragraph{Non-parametric robustness test}

\begin{figure}[t]
  \centering
  \includegraphics[height=3in]{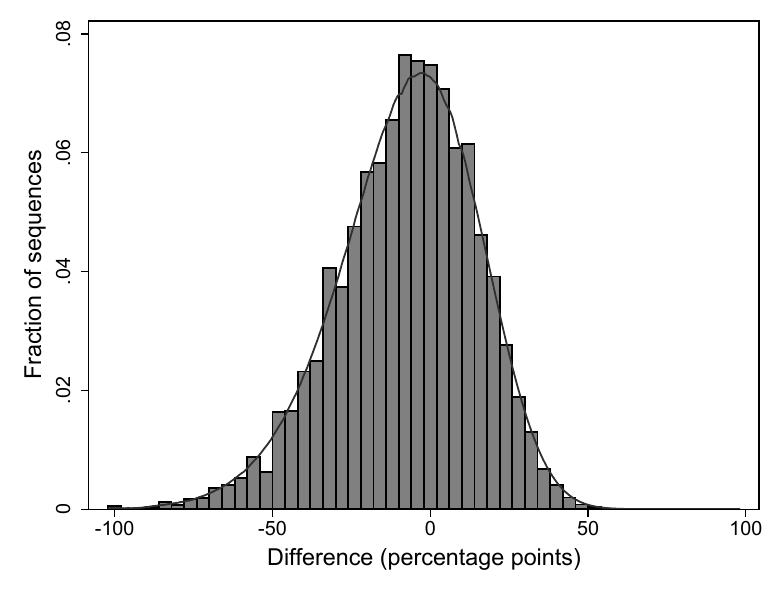}
\caption{The histogram and kernel density plot of the (exact) discrete probability distribution of $\hat{P}^i(ahit|k \ hits) - \hat{P}^i(hit| k\ misses)$, for a single player $i$ with $n=100$ and $n_1=50$, using a variant of the formula for the distribution provided in Web Appendix~\ref{ssec: Appendix recursive diff} (see supplementary materials).\footnotemark \label{fig: Diff Histogram}}
\end{figure}

As a robustness check we perform permutation tests, which are (by construction) invulnerable to the bias. The null hypothesis for a permutation test is that a player is a consistent shooter, i.e. has an i.i.d. fixed (unknown) probability of success. The first step to test for streak shooting in player $i$ is to observe his/her shot sequence and compute the difference in proportions, $\hat{P}^i(hit|k \ hits) - \hat{P}^i(hit| k\ misses)$.  The second step is to compute this difference for each unique rearrangement of the observed sequence; each of these {\it permutations} is equally likely because player $i$'s probability of success is fixed under the null hypothesis.\footnote{Thus, the permutation procedure directly implements GVT's idea of comparing a ``player's performance\ldots to a sequence of hits and misses generated by tossing a coin'' \citep[p. 296]{GilovichValloneTversky--CS--1985}}
The set of unique differences computed in the second step, along with their associated relative frequencies, constitutes the exact sampling distribution of the difference under the null hypothesis (conditional on the observed number of hits). This distribution can then be used for statistical testing (See Appendix~\ref{ssec: Appendix Permutation Details} for details). The distribution is negative-skewed, and can be represented by histograms such as the one shown in Figure~\ref{fig: Diff Histogram},  which provides the \emph{exact} distribution for a player who has hit $50$ out of $100$ shots.

\footnotetext{Uses a bin width of 4 percentage points. The values for the difference are grouped based on the first 6 decimal digits of precision. For this precision, the more than $10^{29}$ distinguishable sequences take on 19,048 distinct values. In the computation of the expected value in Figures~\ref{fig: Bias Prop} and \ref{fig: BiasDiffProp}, each difference is instead represented with the highest floating point precision available.}

Results of the permutation tests agree with those of the bias-corrected tests reported above. In particular, the average difference across shooters
indicates hot hand shooting with a similar level of significance ($p<.01$).\footnote{The procedure in this pooled test involves stratifying the permutations by player. In particular,  we conduct a test of the average of the standardized difference, where for each player the difference is standardized by shifting its mean and scaling its variance under $H_0$. In this case $H_0$: $\prob(\text{success on trial }t\text{ for player }i)=p^i$ for all $t,i$.} Also as before, 5 individual players exhibit significant hot hand shooting ($p<.01$, binomial test).\footnote{As in Footnote~\ref{fn: robust to varying k.}, the results of the permutation test are robust to varying streak length $k$.\label{fn: robust to varying k permutation test}}

\subsection{The hot hand (and bias) in other controlled and semi-controlled studies}\label{ssec: HH in other studies}

A close replication of GVT's controlled shooting experiment is found in \citet{AvugosBar-EliRitovSher--IJSEP--2013}, a study that mimics GVT's design and analysis, but with olympian rather than collegiate players, and fewer shots ($n=40$) per player. From the authors' Table~1 (p. 6), one can derive the average
$\hat{p}(hit|3 \ hits)$ and $\hat{p}(hit| 3\ misses)$ across players, which are roughly .52 and .54, respectively, yielding an average difference in shooting percentages of -2 percentage points.\footnote{We could not analyze the raw data because the authors declined to provide it to us. The data that represents a close replication of GVT is from the betting game phase. Using Table~1, we have $\hat{p}(hit|3 \ hits)=(.56+.52)/2$ and $\hat{p}(hit|3 \ misses)=(.54+.49)/2$, which is the average of the shooting percentage of Group~A in Phase~1 with that of Group~B from Phase~2.} However, Figure~\ref{fig: BiasDiffProp} in Appendix \ref{ssec: Appendix exp diff in prop} shows that the strength of the bias for $n=40$ shots and $p=.5$ (the design target) is -.20. Thus, once the bias is corrected for in this small sample the average difference across shooters becomes roughly +18 percentage points.\footnote{The authors also had another treatment, in which they had shooters rate, before each shot, from 0-100\% on a certainty scale whether they would hit the next shot. If we repeat the analysis on the data from this treatment then the average
$\hat{p}(hit|3 \ hits)$ and $\hat{p}(hit| 3\ misses)$ across players are roughly .56 and .65, respectively, yielding an average difference of -9 percentage points, and a bias-adjusted difference of +11 percentage points.}

\cite{KoehlerConley--JSEP--2003} test for the hot hand in the NBA three point shooting contest, which has been described as an ideal setting in which to study the hot hand \citep{ThalerSunstein2008}. The authors find no evidence of hot hand shooting in their analysis of four years of data. However, as in GVT and \cite{AvugosBar-EliRitovSher--IJSEP--2013}, the conditional probability tests that the authors conduct are vulnerable to the bias.  By contrast, \cite{MillerSanjurjo2015d} collect 28 years of data, which yields 33 players that have taken at least 100 shots; using this dataset, we find that the average bias-corrected difference across players is +8 percentage points ($p<.01$).\footnote{\cite{MillerSanjurjo2015d} also implement the unbiased permutation test procedure of Section \ref{ssec: debiasing GVTs test}.} Further, 8 of the 33 players exhibit significant hot hand shooting ($p<.05$), which itself is statistically significant ($p<.001$, binomial test).

The only other controlled shooting studies that we are aware of are \citet{JagacinskiNewellIssac--JSP--1979} and \citet{MillerSanjurjo--wp--2014}.\footnote{The one exception is a controlled shooting study that involved a single shooter \citet{Wardrop--wp--1999}. After personal communication with the shooter, who conducted the study herself (recording her own shots), we viewed it as not having sufficient control to warrant analysis.}\fnsep\footnote{We thank Tom Gilovich for bringing the study of \citeauthor{JagacinskiNewellIssac--JSP--1979} to our attention. It had gone uncited in the hot hand literature until \cite{MillerSanjurjo--wp--2014}.} Both studies have few shooters (6 and 8, respectively) but many shots across multiple shooting sessions for each player (540 and 900+ shots, respectively). The bias-adjusted average difference in the studies are +7 and +4 percentage points, respectively. In addition, \cite{MillerSanjurjo--wp--2014} find substantial and persistent evidence of hot hand shooting in individual players.\footnote{See \citet{AvugosEtAl--PSE--2013} for a meta-analysis of the hot hand, which includes sports besides basketball. \citet{TverskyGilovich--ASS--2005} argue that evidence for the hot hand in other sports is not relevant to their main conclusion because so long as the hot hand does not exist in basketball, then the perception of the hot hand by fans, players and coaches must necessarily be a cognitive illusion\setcitestyle{square}(see also  \citet{AlterOppenheimer--TR--2006}).\setcitestyle{round}}

Thus, once the bias is accounted for, \emph{conservative} estimates of hot hand effect sizes across all extant controlled and semi-controlled shooting studies are consistently moderate to large.\footnote{The magnitudes of all estimated  effect sizes are conservative for two reasons: (i) if a player's probability of success is not driven merely by feedback from previous shots, but also by other time-varying player (and environment) specific factors, then the act of hitting consecutive shots will serve as only a noisy proxy of the hot state, resulting in measurement error, and an attenuation bias in the estimate (see Appendix~\ref{sec: Appendix Bias adjustment}), and (ii) if the effect of consecutive successes on subsequent success is heterogenous in magnitude (and sign) across players, then an average measure will underestimate how strong the effect can be in certain players.}

\subsection{Belief in the Hot Hand}\label{ssec: Belief in the hot hand}

The results of our reanalysis of GVT's data lead us to a conclusion that is the opposite of theirs: belief in the hot hand is not a cognitive illusion. Nevertheless, it remains possible, perhaps even likely, that professional players and coaches sometimes infer the presence of a hot hand when it does not exist. Similarly, even when in the presence of the hot hand, players may overestimate its influence and respond too strongly to it. By contrast, a hot hand might also go undetected, or be underestimated \citep{StoneArkes--WP--2017}. These questions are important because understanding the extent to which decision makers' beliefs and behavior do not correspond to the actual degree of hot hand shooting could have considerable implications for decision-making more generally.

While GVT's main conclusion was of a binary nature, i.e. based on the question of whether belief in the hot hand is either fallacious or not, they explored hot hand beliefs via a survey of player and coach beliefs, and an incentivized betting task with the Cornell players. In the survey they find that the near universal beliefs in the hot hand do not accord with the lack of hot hand shooting evidence that resulted from their analysis of the shooting data, and in the betting task they found that players were incapable of predicting upcoming shot outcomes successfully, which suggests that even if there were a hot hand, it could not be detected successfully.

However, in light of the results presented in the present paper subjects' responses in GVT's unincentivized survey are actually qualitatively consistent with the evidence presented above.\footnote{See Appendix~B of \citet{MillerSanjurjo--WorkingPaper--2017} for details.} More substantively, GVT's statistical analysis of betting data has recently been shown to be considerably underpowered, as the authors conduct many separate individual bettor level tests rather than pooling the data across bettors \citep{MillerSanjurjo--WorkingPaper--2017}. In addition, GVT misinterpret their measures of bettors' ability to predict. In light of these limitations, \citet{MillerSanjurjo--WorkingPaper--2017} reanalyze GVT's betting data, and find that players on average shoot around +7 percentage points higher when bettors have predicted that the shot will be a hit, rather than a miss ($p<.001$).  This increase is comparable in magnitude to an NBA shooter going from slightly above average to elite in three point percentage.\footnote{ESPN, ``NBA Player 3-Point Shooting Statistics - 2015-16.''   \href{http://www.espn.com/nba/statistics/player/_/stat/3-points}{ http://www.espn.com/nba/statistics/player/\_/stat/3-points} [accessed September 24, 2016].}

\citet{MillerSanjurjo--wp--2014} present complementary evidence on beliefs, in which semi-professional players rank their teammates' respective increases in shooting percentage when on a streak of three hits (relative to their base rates) in a shooting experiment that the rankers do not observe.  Players' rankings are found to be highly correlated with their teammates' actual increases in shooting percentage in this out-of-sample test, yielding an average correlation of  -0.60 ($p < .0001$; where 1 is the rank of the shooter with the perceived largest percentage point increase).

In sum, while it remains possible that professional players' and coaches' hot hand beliefs are poorly calibrated, this claim is not clearly supported by the existing body of evidence.

\section{Conclusion}\label{sec: Conclusion}

We prove the existence of, and quantify, a novel form of selection bias that counter-intuitively arises in some particularly simple analyses of sequential data. A key implication of the bias is that the empirical approach of the canonical hot hand fallacy paper, and its replications, are incorrect. Upon correcting for the bias we find that the data that had previously been interpreted as demonstrating that belief in the hot hand is a fallacy, instead provides substantial evidence that it is not a fallacy to believe in the hot hand.

\singlespace

\bibliographystyle{ecta}

\bibliography{BibliographyCZ,books}

\setstretch{1.25}

\appendix

\clearpage
%\pagenumbering{arabic}
\section{Appendix: Section~\ref{sec: Bias Section} Proofs}\label{sec: Appendix general Proofs}

\subsection{Proof of Theorem~\ref{thm: General bias} (Section~\ref{sec: Bias Section})}\label{ssec: General Proof}

Define $F:=\{\bm{x}\in\{0,1\}^n: I_{k}(\bm{x})\neq\emptyset\}$ to be the sample space of sequences for which $\hat{P}_{k}(\bm{X})$ is well defined. The probability distribution over $F$ is given by $\prob(A |  F):=\prob(A\cap F )/\prob(F)$ for $A\subseteq \{0,1\}^n$, where $\mathbb{P}(\bm{X}=\bm{x})=p^{\sum_{i=1}^n x_i}(1-p)^{n-\sum_{i=1}^n x_i}$.

Let the random variable $X_{\tau}$ represent the outcome of the randomly ``drawn'' trial $\tau$, which is selected as a result of the two-stage procedure that: (i) draws a sequence $\bm{x}$ at random
from $F$, according to the distribution $\prob(\bm{X}=\bm{x}|F)$, and (ii) draws a trial $\tau$ at random from $\{k+1,\dots,n\}$, according to the distribution $\prob(\tau=t|\bm{X}=\bm{x})$. Let
$\tau$ be a uniform draw from the trials in sequence $\bm{X}$ that immediately follow $k$ consecutive successes, i.e.  for $\bm{x}\in F$, $\prob(\tau=t|\bm{X}=\bm{x})=1/|I_{k}(\bm{x})|$ for
$t\in I_{k}(\bm{x})$, and $\prob(\tau=t|\bm{X}=\bm{x})=0$ for $t\in I_{k}(\bm{x})^C\cap\{k+1,\dots,n\}$.\footnote{For $\bm{x}\in F^C$ no trial is drawn, which we can represent as
$\prob(\tau=1|\bm{X}=\bm{x})=1$ (for example).} It follows that the unconditional probability distribution of $\tau$ over all trials that can possibly follow $k$ consecutive successes is given by
$\prob(\tau=t| F)=\sum_{\bm{x}\in F} \prob(\tau=t|\bm{X}=\bm{x},F)\prob(\bm{X}=\bm{x}| F)$, for $t\in \{k+1,\dots,n\}$.  The probability that this randomly drawn trial is a success,
$\prob(X_{\tau}=1|F)$, must be equal to the expected proportion, $E[ \hat{P}_{k}(\bm{X}) |  F ]$.\footnote{The identity follows by the law of total probability, with the key observation that
$\hat{P}_{k}(\bm{x})=\sum_{t\in I_{k}(\bm{x})} x_t\cdot\frac{1}{|I_{k}(\bm{x})|}=\sum_{t=k+1}^n\prob(X_t=1| \tau=t, \bm{X}=\bm{x}, F)\prob(\tau=t|\bm{X}=\bm{x}, F)$.}

Note that $\prob(X_{\tau}=1| F) =\sum_{t=k+1}^n\prob(X_t=1| \tau=t, F)\prob(\tau=t| F)$, and $\prob(\tau=t| F)>0$ for $t\in\{k+1,\dots,n\}$.
Below, we demonstrate that $\prob(X_t=1| \tau=t, F)<p$ when $t<n$, and that $\prob(X_t=1| \tau=n, F)=p$, which, taken together, guarantee that $\prob(X_{\tau}=1| F)<p$.

First we observe that $\prob(X_t=1| \tau=t, F)=\prob(X_t=1| \tau=t,F_t)$, where $F_t:=\{\bm{x}\in\{0,1\}^n: {\textstyle\prod_{i=t-k}^{t-1}} x_i=1 \}$.  Bayes Rule then yields:
\begin{align*}
\frac{\prob(X_t=1| \tau=t,F_t)}{\prob(X_t=0| \tau=t,F_t)}&=\frac{\prob\left(\tau=t\left| X_t=1, F_t \right.\right)}{\prob\left(\tau=t\left| X_t=0, F_t \right.\right)}\frac{\prob(X_t=1|F_t)}{\prob(X_t=0|F_t)}\\
&=\frac{\prob\left(\tau=t\left| X_t=1, F_t \right.\right)}{\prob\left(\tau=t\left| X_t=0, F_t \right.\right)}\frac{p}{1-p}.
\end{align*}

Therefore, for the case of $t\in\{k+1,\dots,n-1\}$, in order to show that  $\prob(X_t=1| \tau=t, F)=\prob(X_t=1| \tau=t,F_t)<p$  it suffices to show that $\prob(\tau=t| X_t=1, F_t)<\prob(\tau=t| X_t=0, F_t)$, which follows below:

\begin{align}
\prob(\tau=t| X_t=0, F_t)=&\sum_{\bm{x}\in F_t : x_t =0} \prob(\tau=t| X_t=0, \bm{X}=\bm{x},  F_t )\prob(\bm{X}=\bm{x} | X_t=0, F_t)\nonumber\\
&=\sum_{\bm{x}\in F_t : x_t =0} \prob(\tau=t| X_t=0, \bm{X}_{-t}=\bm{x}_{-t},  F_t )\prob(\bm{X}_{-t}=\bm{x}_{-t} | X_t=0, F_t)\label{proof: likelihood 3}\\
&>\sum_{\bm{x}\in F_t : x_t =0} \prob(\tau=t| X_t=1, \bm{X}_{-t}=\bm{x}_{-t},  F_t )\prob(\bm{X}_{-t}=\bm{x}_{-t} | X_t=1, F_t) \label{proof: likelihood 4}\\
=&\sum_{\bm{x}\in F_t : x_t=1} \prob(\tau=t| X_t=1, \bm{X}=\bm{x}, F_t )\prob(\bm{X}=\bm{x}| X_t=1, F_t)\nonumber \\
=&\prob(\tau=t| X_t=1, F_t)\nonumber
\end{align}
where in (\ref{proof: likelihood 3}), given $\bm{x}$, we define $\bm{x}_{-t}:=(x_1,\dots,x_{t-1},x_{t+1},\dots, x_n)$.
To obtain the inequality in (\ref{proof: likelihood 4}) we observe that: (i)  $\prob(\bm{X}_{-t}=\bm{x}_{-t} | X_t=0, F_t)=\prob(\bm{X}_{-t}=\bm{x}_{-t} | X_t=1, F_t)$ because $\bm{X}$ is a sequence of i.i.d. Bernoulli trials, and (ii) $\prob(\tau=t |  X_t=1, \bm{X}_{-t}=\bm{x}_{-t},  F_t )<\prob(\tau=t | X_t=0, \bm{X}_{-t}=\bm{x}_{-t},  F_t )$ because $\tau$ is drawn at random (uniformly) from the set $I_k(\bm{x})$, which contains at least one more element (trial $t+1$) if $x_t=1$ rather than $x_t=0$.

For the case of $t=n$ we follow the above steps until (\ref{proof: likelihood 4}), at which point an equality now emerges as $X_n=1$ no longer yields an additional trial from which to draw, because trial $n$ is terminal. This implies that $\prob(\tau=n| X_n=1, F_n)=\prob(\tau=n| X_n=0, F_n)$.

Taking these two facts together: (i) $\prob(X_t=1|\tau=t,F)<p$,    for $k+1\leq t<n$, and (ii) $\prob(X_n=1|\tau=n, F)=p$, it immediately follows that $\prob(X_{\tau}=1| F)<p$.\footnote{Note that the proof does not require that the Bernoulli trials be identically distributed.  Instead, we could allow the probability distribution to vary, with $\prob(X_i=1)=p^i$ for $i=1,\dots,n$, in which case our result would be that  $\prob(X_{\tau}=1| F)<E[p_{\tau}| F]$.}\\
 $\blacksquare$

\subsection{Asymptotic Unbiasedness}\label{ssec: Appendix consistency}

\paragraph{Proof that the proportion is asymptotically unbiased}
To demonstrate that $\hat{P}_{k}(\bm{X})$ is a consistent estimator of $\prob(X_t=1|\prod_{j=t-k}^{t-1} X_j=1)$, first define $Y_{k,i}:=\prod_{j=i-k+1}^{i}X_j$ for $i\geq k$. With this, $\hat{P}_{k}(\bm{X})=\sum_{i=k+1}^n Y_{k+1,i}/\sum_{i=k}^{n-1} Y_{k,i}$.  Note that each of the respective sequences $\{Y_{k,i}\}$, $\{Y_{k+1,i}\}$ are asymptotically uncorrelated ($k$ fixed). Therefore, their time averages converge to their respective means almost surely, i.e. $1/(n-k)\sum_{i=k}^{n-1} Y_{k,i}\stackrel{a.s.}{\longrightarrow}E[Y_{k,i}]=p^k$, and  $1/(n-k)\sum_{i=k+1}^n Y_{k+1,i}\stackrel{a.s.}{\longrightarrow}E[Y_{k+1,i}]=p^{k+1}$.\footnote{See Definition~3.55 and Theorem~3.57 from \citet{White--book--1999}.} The continuous mapping theorem implies that $\hat{P}_{k}(\bm{X})\stackrel{a.s.}{\longrightarrow} p=\prob(X_t=1\ | \ \prod_{j=t-k}^{t-1} X_j=1)$, which in turn implies consistency.

\paragraph{Proof that weighted proportions are asymptotically unbiased}

In order to prove the assertion made in Footnote~\ref{fn: multiple sequences} that the weighted average proportion over multiple realized sequences is a consistent estimator of $\prob(X_t=1|\prod_{j=t-k}^{t-1} X_j=1)$, we first define $Y_{k,i}:=\prod_{j=i-k+1}^{i}X_j$ for $i\geq k$, just as we did in the previous proof. Then, we note that: (i) the number of trials that follow $k$ consecutive successes in the weighted proportion taken over $T$ sequences is given by $\sum_{t=1}^T Z_{k,t}$, where $Z_{k,t}=\sum_{i=n(t-1)+k}^{nt-1} Y_{k,i}$, and (ii) the number of successes on these trials is given by $\sum_{t=1}^T Z_{k+1,t}$, where $Z_{k+1,t}=\sum_{i=n(t-1)+k+1}^{nt} Y_{k+1,i}$. Because $Z_{k,t}$ are i.i.d.  with $E[Z_{k,t}]=(n-k)p^k$, it follows that $ 1/T\sum_{t=1}^T Z_{k,t}\stackrel{a.s.}{\longrightarrow} E[Z_{k,t}]=(n-k)p^k$;  similarly, $ 1/T\sum_{t=1}^T Z_{k+1,t}\stackrel{a.s.}{\longrightarrow} E[Z_{k+1,t}]=(n-k)p^{k+1}$.  Then, the continuous mapping theorem yields the desired consistency of the weighted proportion (after sequence $T$), i.e. $\sum_{t=1}^T Z_{k+1,t}/\sum_{t=1}^T Z_{k,t}\stackrel{a.s.}{\longrightarrow} p=\prob(X_t=1\ | \ \prod_{j=t-k}^{t-1} X_j=1)$.

\subsection{Formula for the expected proportion (special case of $k=1$)}\label{sec: Appendix k=1 Proofs}

The following lemma shows that the expected proportion $\hat{P}_{1}(\bm{X})$, conditional on a known number of successes $N_1(\bm{X})=n_1$, satisfies the sampling-without-replacement formula, which for any given trial is less than the probability of success  $\prob(X_i |N_1(\bm{X})=n_1)=\frac{n_1}{n}$.

\begin{lemma}\label{lemma: E(P11|N1)}
Let $n>1$. Then
\begin{equation}
E\left.\left[\ \hat{P}_{1}(\bm{X})\ \right| \ I_{1}(\bm{X}) \neq \emptyset  ,\ N_1(\bm{X})=n_1 \ \right] = \frac{n_1-1}{n-1} \label{eq: E(P11|N1)}
\end{equation}
 for $0\leq n_1 \leq n$.
 \end{lemma}
\textbf{Proof:}
As in the proof of Theorem~\ref{thm: General bias}, let $\tau$ be drawn at random from $I_{1}(\bm{X})$, which is non-empty when $N_1(\bm{X})=n_1 \geq 2$ (the result is trivial when $n_1=1$). In order to ease notation we let probability $\prob(\cdot)$ represent the conditional probability $\prob(\cdot|N_1(\bm{X})=n_1)$, which is defined over the subsets of $\{\bm{x}\in\{0,1\}^n: N_1(\bm{x})=n_1\}$.
\noindent
\begin{align}
E[ \hat{P}_{1}(\bm{X}) |  N_1(\bm{X}) = n_1, I_{1}(\bm{X})\neq \emptyset ]&=\prob(X_{\tau}=1) \label{eq: Bayes k=1 proof 1}\\
&=\prob(X_{\tau}=1|\tau<n)\prob(\tau<n)+\prob(X_n=1|\tau=n)\prob(\tau=n)\nonumber\\
&=\sum_{t=2}^{n-1}\prob(X_t=1|\tau=t)\frac{1}{n-1}+\prob(X_n=1|\tau=n)\frac{1}{n-1}\label{eq: Bayes k=1 proof 2}\\
&=\frac{n-1}{n-2}\left(\frac{n_1}{n}-\frac{1}{n-1}\right)\frac{n-2}{n-1}+\frac{n_1}{n}\frac{1}{n-1}\label{eq: Bayes k=1 proof 3}\\
&=\frac{n_1-1}{n-1}\nonumber
\end{align}
In (\ref{eq: Bayes k=1 proof 1}), equality follows by an argument analogous to that provided in the proof of Theorem~\ref{thm: General bias}. In (\ref{eq: Bayes k=1 proof 2}), equality follows from the fact that $\prob(\tau=t)=1/(n-1)$ for all $t \in \{2,3,\ldots,n\}$.\footnote{Note that $\prob(\tau=t)=\sum_{\bm{x}: N_1(\bm{x})=n_1} \prob(\tau=t|\bm{X}=\bm{x})\prob(\bm{X}=\bm{x})=\sum_{\bm{x}: N_1(\bm{x})=n_1, x_{t-1}=1} \frac{1}{n_1-x_n}\frac{1}{\binom{n}{n_1}}=\frac{1}{\binom{n}{n_1}}\left[\binom{n-2}{n_1-1}\frac{1}{n_1}+\binom{n-2}{n_1-2}\frac{1}{n_1-1}\right]=\frac{1}{n-1}$.}
In (\ref{eq: Bayes k=1 proof 3}), equality follows from using an application of Bayes rule to derive $\prob(X_t=1|\tau=t)$, which satifies:
\begin{equation}
\prob(X_t=1|\tau=t)=
\begin{cases}
\frac{n-1}{n-2}\left(\frac{n_1}{n}-\frac{1}{n-1}\right) &\mbox{for }  t=2,\dots,n-1 \\
\frac{n_1}{n}  &\mbox{for } t=n
\end{cases}\label{eq: Bayes k=1 likelihood}
\end{equation}
In particular,
\begin{align}
\prob(X_t=1|\tau=t)&=\frac{\prob(\tau =t|X_{t-1}=1,X_t=1)\prob(X_{t-1}=1| X_t=1)\prob(X_t=1)}{\prob(\tau=t)}\nonumber\\
&=\prob(\tau =t|X_{t-1}=1,X_t=1)\frac{n_1(n_1-1)}{n}\label{eq: Bayes in alt proof}
\end{align}
where for all $t$, $\prob(X_{t-1}=1| X_t=1)=(n_1-1)/(n-1)$, which is the likelihood that relates to sampling-without-replacement.  For $t<n$, $\prob(\tau =t|X_{t-1}=1,X_t=1)$, which is the likelihood that relates to the arrangement of successes and failures, satisfies:
\begin{align*}
  \prob(\tau =t|X_{t-1}=1,X_t=1)&=E\left.\left[\ \frac{1}{M}\ \right|\ X_{t-1}=1, X_t=1\right]\\
  &=\sum_{x\in\{0,1\}}E\left.\left[\ \frac{1}{M}\ \right|\ X_{t-1}=1, X_t=1, X_n=x\right]\prob(X_n=x|X_{t-1}=1, X_t=1)\\
   &=\frac{1}{n_1}\frac{n_0}{n-2} + \frac{1}{n_1-1}\frac{n_1-2}{n-2}\\
      &=\frac{1}{n-2}\left(\frac{n_0}{n_1}+ \frac{n_1-2}{n_1-1}\right)
\end{align*}
where $M:=|I_{1}(\bm{X})|$, i.e. $M=n_1-X_n$. In the case that $t=n$, clearly $\prob(\tau =n|X_{n-1}=1,X_n=1)=\frac{1}{n_1-1}$.\\
 $\blacksquare$

\paragraph{Formulae for expected value of the proportion}

The conditional expectation in Lemma~\ref{lemma: E(P11|N1)} can be combined with $\prob(N_1(\bm{X})=n_1| I_{1}(\bm{X}) \neq \emptyset )$  to express the expected proportion in terms of just $n$ and $p$.\footnote{In a comment written about this paper, \citet{RinottBar-Hillel--WP--2015} provide an alternative proof for this theorem.}

\begin{theorem}\label{thm: E(P11)}
Let $n>2$ and $0<p<1$. Then
\begin{equation}
E\left.\left[\ \hat{P}_{1}(\bm{X}) \ \right| \  I_{1}(\bm{X}) \neq \emptyset     \ \right]=\frac{\left[p-\frac{1-(1-p)^n}{n}\right]\frac{n}{n-1}}{1-(1-p)^{n-1}}<p\label{eq: E(P11) proof}
\end{equation}
\end{theorem}
\textbf{Proof:} We first observe that in light of Lemma~\ref{lemma: E(P11|N1)}, Equation~\ref{eq: E(P11) proof} can be written as follows:
\begin{align*}
E\left.\left[\ \hat{P}_{1}(\bm{X}) \ \right| \ I_{1}(\bm{X})\neq \emptyset     \ \right] &=E\left[   E\left.\left[\ \hat{P}_{1}(\bm{X})\ \right| \ I_{1}(\bm{X})\neq \emptyset ,\ N_1(\bm{X})=n_1 \ \right]\right]\\
&=E\left.\left[  \frac{N_1(\bm{x})-1}{n-1}\right| I_{1}(\bm{X})\neq \emptyset\  \right]
\end{align*}
The expected value can then be computed using the binomial distribution, which yields:
\begin{align*}
E\left.\left[  \frac{N_1(\bm{x})-1}{n-1}\right| I_{1}(\bm{X})\neq \emptyset\  \right]&=C \sum_{n_1=1}^n  p^{n_1}(1-p)^{n-n_1}\left[\binom{n}{n_1}- U(n,n_1) \right]\cdot  \frac{n_1-1}{n-1} \\
&=\frac{\sum_{n_1=2}^n\binom{n}{n_1}p^{n_1}(1-p)^{n-n_1}\frac{n_1-1}{n-1}}{1-(1-p)^n-p(1-p)^{n-1}}\\
&=\frac{\frac{1}{n-1}\left[\left(np-np(1-p)^{n-1}\right)-\left(1-(1-p)^n-np(1-p)^{n-1}\right)\right]}{1-(1-p)^n-p(1-p)^{n-1}}\\
&=\frac{\left[p-\frac{1-(1-p)^n}{n}\right]\frac{n}{n-1}}{1-(1-p)^{n-1}}
\end{align*}
where $U(n,n_1)$ is the number of sequences with $n_1$ successes for which the proportion is undefined, and $C$ is the constant that normalizes the total probability to 1. The second line follows because $U_{1}(n,n_1)=0$ for $n_1>1$, $U_{1}(n,0)=U_{1}(n,1)=1$, and $C=1/[1-(1-p)^n-p(1-p)^{n-1}]$.

Finally, by letting $q:=1-p$ it is straightforward to show that the bias in $\hat{P}_{1}(\bm{X})$ is negative:
\begin{align*}
E\left.\left[\ \hat{P}_{1}(\bm{X})-p \ \right| \ I_{1}(\bm{X})\neq \emptyset     \ \right]
&=\frac{\left[p-\frac{1-q^n}{n}\right]\frac{n}{n-1}}{1-q^{n-1}}-p\\
&=\frac{(n-1)(q^{n-1}-q^n)-(q-q^n)}{(n-1)(1-q^{n-1})}\\
&<0
\end{align*}
The inequality follows from $f(x)=q^x$ being strictly decreasing and convex, which implies that $q-q^n>(n-1)(q^{n-1}-q^n)$.\\
 $\blacksquare$

\subsection{Expected difference in proportions}\label{ssec: Appendix exp diff in prop}
Let $D_k$ be the difference in the probability of success when comparing trials that immediately follow $k$ consecutive successes with trials that immediately follow $k$ consecutive failures. That is, $D_k:=\prob(X_t=1\ | \ \prod_{j=t-k}^{t-1} X_j=1)-\prob(X_t=1\ | \ \prod_{j=t-k}^{t-1} (1-X_j)=1)$. An estimator of $D_k$ that is used in the hot hand fallacy literature (see Section~\ref{sec: HF}) is $\hat{D}_k(\bm{x}):=\hat{P}_k(\bm{x})-[1-\hat{Q}_k(\bm{X})]$, where $\hat{Q}_k(\bm{X})$ is the proportion of failures on the subset of trials that immediately follow $k$ consecutive failures, $J_{k}(\bm{X}):=\{ j : {\textstyle\prod_{i=j-k}^{j-1}} (1-X_i) =1\}\subseteq \{k+1,\dots,n\}$.

\subsubsection{Proof of the bias in the difference}

We extend the proof of Theorem \ref{thm: General bias} to show that $\hat{D}_k(\bm{X})$ is a biased estimator of $D_k$. Recall that $I_{k}(\bm{X})$ is the subset of trials  that immediately follow $k$ consecutive successes, i.e. $I_{k}(\bm{X}):=\{ i: {\textstyle\prod_{j=i-k}^{i-1}} X_j =1\}\subseteq \{k+1,\dots,n\}$. Analogously, let $J_{k}(\bm{X})$ be the subset of trials  that immediately follow $k$ consecutive failures, i.e. $J_{k}(\bm{X}):=\{ j : {\textstyle\prod_{i=j-k}^{j-1}} (1-X_i) =1\}\subseteq \{k+1,\dots,n\}$.

\begin{theorem}\label{thm: General bias diff}
Let $\bm{X}=\{X_i\}_{i=1}^n$, $n\geq 3$, be a sequence of independent Bernoulli trials, each with probability of success $0<p<1$. Let $\hat{P}_{k}(\bm{X})$ be the proportion of successes on the subset of trials $I_{k}(\bm{X})$ that immediately follow $k$ consecutive successes, and $\hat{Q}_k(\bm{X})$ be the proportion of failures on the subset of trials $J_{k}(\bm{X})$ that immediately follow $k$ consecutive failures. $\hat{D}_k(\bm{x}):=\hat{P}_k(\bm{x})-[1-\hat{Q}_k(\bm{X})]$ is a biased estimator of $D_k:=\prob(X_t=1\ | \ \prod_{j=t-k}^{t-1} X_j=1)-\prob(X_t=1\ | \ \prod_{j=t-k}^{t-1} (1-X_j)=1)\equiv 0$ for all $k$ such that $1\leq k< n/2$. In particular,

\begin{equation}
E[\ \hat{D}_k(\bm{X}) \ | \ I_{k}(\bm{X}) \neq \emptyset, J_{k}(\bm{X}) \neq \emptyset    \ ]< 0
\end{equation}
\end{theorem}
\textbf{Proof:} Following the notation used in the proof of Theorem~\ref{thm: General bias}, let $F:=\{\bm{x}\in\{0,1\}^n: I_{k}(\bm{x})\neq\emptyset\}$ and $G:=\{\bm{x}\in\{0,1\}^n: J_{k}(\bm{x})\neq\emptyset\}$. We will show the following:

\begin{align}
E[\hat{D}_k(\bm{x})|F, G ]&= E[\hat{P}_{k}(\bm{X})|F, G]+E[\hat{Q}_{k}(\bm{X})|F, G]-1\notag\\
&=\prob(X_{\tau}=1|F, G)+\prob(X_{\sigma}=0|F, G)-1\label{eq: Diff them appendix 1}\\
&<p+(1-p)-1  \label{eq: Diff them appendix 2}\\
&=0 \label{eq: Diff them appendix 3}
\end{align}
where in (\ref{eq: Diff them appendix 1}), as in the proof of Theorem~\ref{thm: General bias}, $\tau$ is a random draw from $I_{k}(\bm{x})$ and $\sigma$ is an analogous random draw from $J_{k}(\bm{x})$.  In particular, we will demonstrate that the inequality in
(\ref{eq: Diff them appendix 2}) holds by showing that $\prob(X_\tau=1|F, G)<p$, which, by symmetry, implies that $\prob(X_\sigma=0|F, G)<1-p$.

To show that $\prob(X_\tau=1|F, G)<p$ we use an approach similar to that presented in the proof of Theorem~\ref{thm: General bias}. In particular, note that $\prob(X_{\tau}=1| F,G) =\sum_{t=k+1}^n\prob(X_t=1| \tau=t, F,G)\prob(\tau=t| F,G)$, and $\prob(\tau=t| F,G)>0$ for $t\in\{k+1,\dots,n\}$.
In what follows, we demonstrate that $\prob(X_t=1| \tau=t, F,G)<p$ when $t<n$, and that $\prob(X_t=1| \tau=n, F,G)=p$, which, taken together, guarantee that $\prob(X_{\tau}=1| F,G)<p$.

First we observe that $\prob(X_t=1| \tau=t, F,G)=\prob(X_t=1| \tau=t,F_t,G)$, where $F_t:=\{\bm{x}\in\{0,1\}^n: {\textstyle\prod_{i=t-k}^{t-1}} x_i=1 \}$.  Bayes Rule then yields:
\begin{align*}
\frac{\prob(X_t=1| \tau=t,F_t,G)}{\prob(X_t=0| \tau=t,F_t,G)}&=\frac{\prob\left(\tau=t,G\left| X_t=1, F_t \right.\right)}{\prob\left(\tau=t,G\left| X_t=0, F_t \right.\right)}\frac{\prob(X_t=1|F_t)}{\prob(X_t=0|F_t)}\\
&=\frac{\prob\left(\tau=t,G\left| X_t=1, F_t \right.\right)}{\prob\left(\tau=t,G\left| X_t=0, F_t \right.\right)}\frac{p}{1-p}.
\end{align*}
Therefore, for the case of $t\in\{k+1,\dots,n-1\}$, in order to show that  $\prob(X_t=1| \tau=t, F,G)=\prob(X_t=1| \tau=t,F_t,G)<p$   it  suffices to show that $\prob(\tau=t,G| X_t=1, F_t)<\prob(\tau=t,G| X_t=0, F_t)$, which follows below:
\begin{align}
\prob(\tau=t,G| X_t=0, F_t)=&\sum_{\substack{\bm{x}\in F_t\cap G: \\ x_t =0}} \prob(\tau=t, \bm{X}=\bm{x} | X_t=0,  F_t )\nonumber\\
=&\sum_{\substack{\bm{x}\in F_t\cap G: \\ x_t =0\\(1,\bm{x}_{-t})\in F_t\cap G}} \prob(\tau=t, \bm{X}=\bm{x} | X_t=0,  F_t )\label{proof: diff likelihood 1}\\
&\quad+\sum_{\substack{\bm{x}\in F_t\cap G: \\ x_t =0\\(1,\bm{x}_{-t})\notin F_t\cap G}} \prob(\tau=t, \bm{X}=\bm{x} | X_t=0,  F_t )\nonumber\\
\geq&\sum_{\substack{\bm{x}\in F_t\cap G: \\ x_t =0\\(1,\bm{x}_{-t})\in F_t\cap G}} \prob(\tau=t, \bm{X}=\bm{x} | X_t=0,  F_t )\nonumber\\
=&\sum_{\substack{\bm{x}\in F_t\cap G: \\ x_t =0\\(1,\bm{x}_{-t})\in F_t\cap G}} \prob(\tau=t | \bm{X}=\bm{x}, X_t=0,  F_t )\prob(\bm{X}=\bm{x} | X_t=0, F_t)\nonumber\\
=&\sum_{\substack{\bm{x}\in F_t\cap G: \\ x_t =0\\(1,\bm{x}_{-t})\in F_t\cap G}} \prob(\tau=t |  X_t=0, \bm{X}_{-t}=\bm{x}_{-t},  F_t )\prob(\bm{X}_{-t}=\bm{x}_{-t} | X_t=0, F_t)\nonumber\\
>&\sum_{\substack{\bm{x}\in F_t\cap G: \\ x_t =0\\(1,\bm{x}_{-t})\in F_t\cap G}} \prob(\tau=t |  X_t=1,\bm{X}_{-t}=\bm{x}_{-t},  F_t )\prob(\bm{X}_{-t}=\bm{x}_{-t} | X_t=1, F_t)\label{proof: diff likelihood 3}\\
=&\sum_{\substack{\bm{x}\in F_t\cap G: \\ x_t =1\\(1,\bm{x}_{-t})\in F_t\cap G}} \prob(\tau=t | \bm{X}=\bm{x}, X_t=1,  F_t )\prob(\bm{X}=\bm{x} |X_t=1,  F_t)\nonumber\\
=&\sum_{\substack{\bm{x}\in F_t\cap G: \\ x_t =1}} \prob(\tau=t,\bm{X}=\bm{x} |  X_t=1,  F_t )\nonumber\\
=&\prob(\tau=t,G| X_t=1, F_t)\nonumber
\end{align}
where in (\ref{proof: diff likelihood 1}), given $\bm{x}$, we define $\bm{x}_{-t}:=(x_1,\dots,x_{t-1},x_{t+1},\dots, x_n)$, and $(b,\bm{x}_{-t}) :=  (x_1,\dots,x_{t-1},b,x_{t+1}, \dots, x_n)$.\footnote{Note that the second sum will have no terms for $t\geq n-k$.} The inequality in (\ref{proof: diff likelihood 3}) follows for the same reason as the inequality in line (\ref{proof: likelihood 4}) of Theorem~\ref{thm: General bias}. In particular, $\prob(\bm{X}_{-t}=\bm{x}_{-t}| X_t=0, F_t)=\prob(\bm{X}_{-t}=\bm{x}_{-t}| X_t=1, F_t)$ because $\bm{X}$ is a sequence of i.i.d. Bernoulli trials, and  $\prob(\tau=t |  X_t=1, \bm{X}_{-t}=\bm{x}_{-t},  F_t )<\prob(\tau=t |  X_t=0,\bm{X}_{-t}=\bm{x}_{-t},  F_t )$ because $\tau$ is drawn at random (uniformly) from the set $I_k(\bm{x})$,
which contains at least one more element (trial $t+1$) if $x_t=1$ rather than $x_t=0$.

For the case of $t=n$ we follow the above steps until (\ref{proof: diff likelihood 3}), at which point an equality now emerges, as $X_n=1$ no longer  yields an additional trial from which to draw, because trial $n$ is terminal. This implies that $\prob(\tau=n| X_n=1, F_n, G)=\prob(\tau=n| X_n=0, F_n,G)$.

Taking these two facts together: (i) $\prob(X_t=1|\tau=t,F,G)<p$,    for $k+1\leq t<n$, and (ii) $\prob(X_n=1|\tau=n,F,G)=p$, it immediately follows that $\prob(X_{\tau}=1| F,G)<p$.\\
 $\blacksquare$

\subsubsection{Formula for the expected difference in proportions (special case of $k=1$)}
In the case of $k=1$ the expected difference in proportions admits a simple representation that is independent of $p$.

\begin{theorem}\label{thm: E(D1)}
Let $\hat{D}_1(\bm{X}):=\hat{P}_{1}(\bm{X})-(1-\hat{Q}_{1}(\bm{X}))$. If $n>2$ and $0<p<1$ then
\[
E\left.\left[\ \hat{D}_1(\bm{X}) \ \right| \ I_{1}(\bm{X}) \neq \emptyset, J_{1}(\bm{X}) \neq \emptyset    \ \right] = -\frac{1}{n-1}
\]

\end{theorem}
\textbf{Proof:} The method of proof is to first show that if $n>2$ and  $1\leq n_1\leq n-1$ then:
\[
E\left.\left[\ \hat{D}_1(\bm{X}) \ \right| \ N_1(\bm{X}) = n_1, I_{1}(\bm{X})\neq \emptyset, J_{1}(\bm{X})\neq \emptyset     \ \right] = -\frac{1}{n-1}
\]
which leaves us just one step from the desired result.

First, consider the case that $1<n_1<n-1$. In this case $\hat{D}_1(\bm{x}):=\hat{P}_{1}(\bm{x})-(1-\hat{Q}_{1}(\bm{x}))$ is defined for all sequences. Therefore, by the linearity of the expectation, and applying Lemma~\ref{lemma: E(P11|N1)} to $\hat{Q}_{1}(\bm{X})$ (by symmetry), we have:
 \begin{align*}
 E[ \hat{D}_1(\bm{X}) | N_1(\bm{X}) = n_1]=& E[\hat{P}_{1}(\bm{X}) |N_1(\bm{X}) = n_1 ]-E(1-\hat{Q}_{1}(\bm{X}) | N_1(\bm{X}) = n_1]\\
 =&\frac{n_1-1}{n-1}-\left(1-\frac{n_0-1}{n-1}\right)\\
 =&-\frac{1}{n-1}
\end{align*}
If $n_1=1$ then $\hat{D}_1$ is defined for all sequences that do not have a $1$ in the final position; there are $n-1$ such sequences.  The sequence with the $1$ in the first position yields $\hat{D}_1=0$, while the other $n-2$ sequences yield $\hat{D}_1=-1/(n-2)$. Therefore, $ E\left.\left[\ \hat{D}_1(\bm{X}) \ \right| \ N_1(\bm{X}) = 1, I_{1}(\bm{X})\neq \emptyset, J_{1}(\bm{X})\neq \emptyset        \ \right]=-1/(n-1)$.

Now consider the case of $n_1=n-1$. The argument for this case is analogous, with $\hat{D}_1$ undefined for the sequence with the zero in the last position, equal to 0 for the sequence with the zero in the first position, and equal to $-1/(n-2)$ for all other sequences.

Finally, that the conditional expectation is independent of $N_1(\bm{x})$ implies that $E[D_{1}(\bm{X})\ | \ I_{1}(\bm{X})\neq \emptyset, J_{1}(\bm{X})\neq \emptyset     \ ]$ is independent of $p$, yielding the result.\\ $\blacksquare$

\subsubsection{Quantifying the bias for the difference}

\begin{figure}[t]
\begin{centering}
  \includegraphics[height=4in]{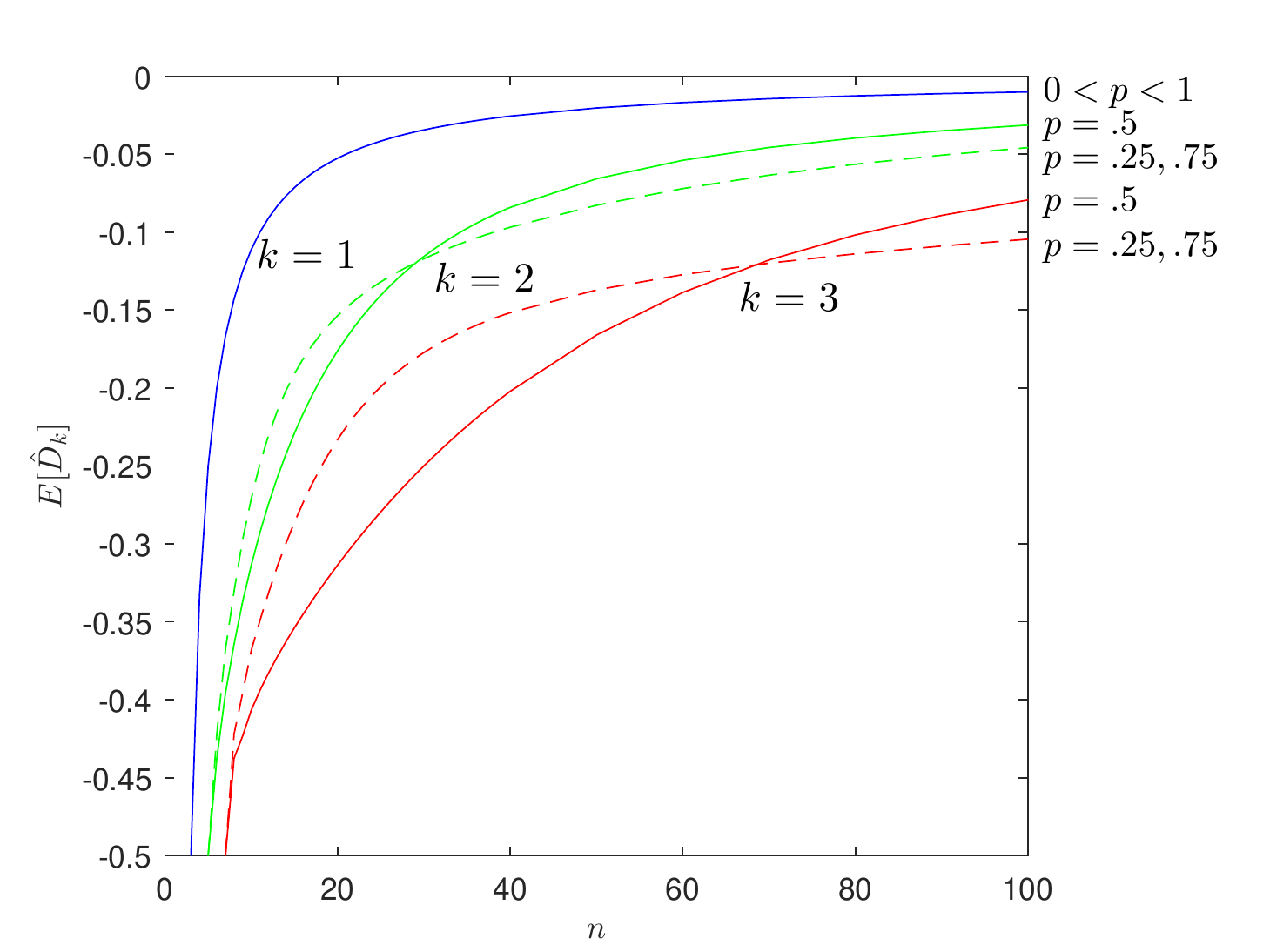}
  \caption{For the expected difference in the proportion of successes, as a function of $n$, three values of $k$, and various probabilities of success $p$, using the formula provided in Web Appendix~\ref{ssec: Appendix recursive diff}.}
  \label{fig: BiasDiffProp}
\end{centering}
\end{figure}
%fig-BiasDiffProp 

Figure~\ref{fig: BiasDiffProp} contains a plot of $E[\ \hat{D}_k(\bm{X}) \ | \ I_{k}(\bm{X}) \neq \emptyset, J_{k}(\bm{X}) \neq \emptyset    \ ]$ as a function of the number of trials $n$, and for $k=1,2,$ and $3$.  Because the bias is dependent on $p$ when $k>1$, the difference is plotted for various values of $p$. These expected differences are obtained from the formula provided in Web Appendix~\ref{ssec: Appendix recursive diff}. The magnitude of the bias is simply the absolute value of the expected difference. As with the bias in the proportion (see Figure \ref{fig: Bias Prop}), the bias in the difference is substantial even when $n$ is relatively large.

\clearpage
\section{Appendix: Size of the bias when the DGP is hot hand/streak shooting}\label{sec: Appendix Bias adjustment}

In Section~\ref{ssec: debiasing GVTs test} the correction to GVT's estimate of the hot hand effect (and test statistic)  is based on the magnitude of the bias under the assumption that the shooter has a fixed probability of success (Bernoulli process).  However, if the underlying data generating process (DGP) instead represents hot hand or streak shooting, then the size of the bias changes.  While many DGPs can produce hot hand shooting, arguably the most natural  are those discussed in \citet{GilovichValloneTversky--CS--1985}, as they reflect lay conceptions of the hot hand and streak shooting.  While GVT take no particular stance on which lay definition is most appropriate, they do identify hot hand and streak shooting with: (i) ``non-stationarity'' (the zone, flow, in the groove, in rhythm), and (ii) ``positive association'' (success breeds success).  We label (i) as a {\it regime shift} model, and interpret it as capturing the idea that a player's probability of success may increase due to some factor that is unrelated to previous outcomes, so unobservable to the econometrician.  This can be modeled naturally as a hidden markov chain over the player's (hidden) ability state.  We label (ii) as a {\it positive feedback} model, because it can be interpreted as capturing the idea that positive feedback from a player's previous shot outcomes can affect his/her subsequent probability of success. This can be modeled naturally as an autoregressive process, or equivalently as a markov chain over shot outcomes.\footnote{A positive feedback model need not be stationary.}

\begin{figure}[t]
    \begin{centering}
      \includegraphics[height=2.2 in]{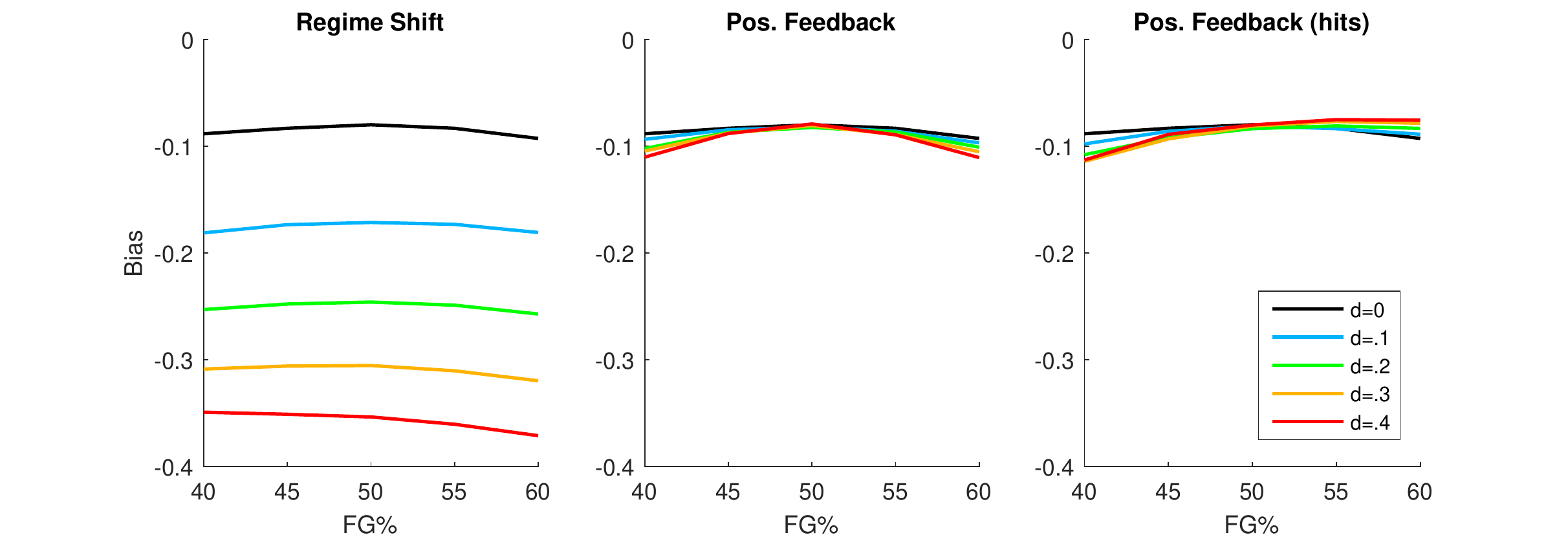}
      \caption{The bias for three types of hot hand and streak shooting data generating processes (DGPs), where $FG\%$ is the expected overall field goal percentage from the DGP, and $d$ represents the change in the player's underlying probability of success. When $d=0$ each model reduces to a Bernoulli process. Therefore the black line represents the bias in a Bernoulli proccess ($n=100$ trials, $k=3$).}
      \label{fig: BiasRobustDGP}
\end{centering}
\end{figure}

In Figure~\ref{fig: BiasRobustDGP} we plot the bias in the estimator  of the difference in probability of success when on a hit streak rather than miss streak, $\hat{D}_3$, for three alternative DGPs, each of which admits the Bernoulli process as a special case.\footnote{Each point is the output of a simulation with 10,000 repetitions of 100 trials from the DGP.} The first panel corresponds to the ``regime shift'' DGP in which the difference in the probability of success between the ``hot'' state and the ``normal'' state is given by $d$ (where $d=0$ represents Bernoulli shooting),\footnote{In particular, let $Q$ be the hidden markov chain over the ``normal'' state ($n$) and the ``hot'' state ($h$), where the probability of success in the normal state is given by $p_n$, and the probability of success in the hot state is given by $p_h$, with the shift in probability of success given by $d:=p_h-p_n$
    \[Q:=\left(
        \begin{array}{cc}
          q_{nn} & q_{nh} \\
          q_{hn} & q_{hh} \\
        \end{array}
      \right)\]
    Where $q_{nn}$ represents the probability of staying in the ``normal'' state, and $q_{nh}$ represents the probability of transitioning from the ``normal'' to the ``hot'' state, etc.  Letting $\pi=(\pi_n,\pi_h)$ be the stationary distribution, we find that the magnitude of the bias is not very sensitive to variation in the stationary distribution and transition probabilities within a plausible range (i.e. $\pi_h\in[.05,.2]$ and $q_{hh}\in(.8,.98)$), while it varies greatly with the difference in probabilities $d$ and the overall expected shooting percentage $p=p_n+\pi_h d$.  In the graph, for each $d$ and $p$ (FG\%), we average across values for the stationary distribution ($\pi_h$) and transition probability ($q_{hh}$).}   the second panel corresponds to the ``positive feedback'' DGP in which hitting (missing) $3$ shots in a row increases (decreases) the probability of success by $d/2$, and the third panel corresponds to the ``positive feedback (for hits)'' DGP in which positive feedback operates for hits only, making the probability of success increase by $d$ whenever $3$ hits in a row occurs.  Within each panel of the figure, the bias, which is the expected difference between $\hat{D}_3$, the estimator of the shift in the probability of success, and $d$, the true shift in the probability of success, is depicted as a function of the expected overall shooting percentages (from 40 percent to 60 percent), for four true shifts in the underlying probability ($d\in\{.1,.2,.3,.4\}$).\footnote{Results are similar if the DGP instead has negative feedback, i.e. $d\in\{-.1,-.2,-.3,-.4\}$.}

  Observe that when the true DGP is a player with a hot hand, the bias is typically more severe, or far more severe, than the bias associated with a Bernoulli DGP.  In particular, the bias in the ``regime shift'' model is particularly severe, which arises from two sources: (i) the bias discussed in Section~\ref{sec: Bias Section}, and (ii) an attenuation bias, due to measurement error, as hitting 3 shots in a row is an imperfect proxy for the ``hot state.''\footnote{In practice observers may have more information than the econometrician (e.g. shooting mechanics, perceived confidence, or lack thereof, etc.), so may be subject to less measurement error.}  The bias in the positive feedback DGP is uniformly below the bias for a Bernoulli shooter.  For the DGP in which positive feedback operates only for hits, the bias is stronger than that of Bernoulli shooters for expected shooting percentages below 50 percent (as in GVTs data), and slightly less strong for shooting percentage above 50 percent.  As the true DGP is likely some combination of a regime shift and positive feedback, it is reasonable to conclude that the empirical approach in Section~\ref{ssec: debiasing GVTs test} should be expected to (greatly) understate the true magnitude of any underlying hot hand.

  The intuition for why the introduction of regime shift elements increases the strength of the bias so considerably is that if a player's probability of success is not driven merely by feedback from previous shots, but also by other time-varying player (and environment) specific factors, then the act of hitting consecutive shots will serve as only a noisy proxy of the hot state, resulting in measurement error, and an attenuation bias in the estimate.  This type of measurement error is similar to what \citet{Stone--AS--2012} uncovered in the relationship between autocorrelation in outcomes and autocorrelation in ability when considering a DGP that contains autocorrelation in ability.
\clearpage
\section{Appendix: Additional analyses, and details for Section~\ref{sec: HF} \label{sec: Appendix more from sec 3} }

\subsection{An alternative (pooled) analysis of shooting data \label{ssec: Appendix Pooled analysis}}
An alternative approach to testing for streak shooting across players is to pool all shots from the ``3 hits'' and ``3 misses'' categories (discarding the rest), then use a linear probability model to estimate the effect of a shot falling in the ``3-hits'' category. If the implementation of GVT's design met the goal of placing each player in a position in which his or her probability of success is .5, then this approach would be analogous to re-weighting the under-weighted coin flips in Table~\ref{tab: bias} of Section~\ref{sec: Introduction}. With 2515 shots, the bias is minimal and the estimate in this case is +17 percentage points ($p<.01$, $S.E.=3.7$).  Because GVT's design goal is difficult to implement in practice, this approach will introduce an upward bias, due to aggregation, if the  probability of success varies across players.  Adding fixed effects in a regression will control for this aggregation bias, but strengthens the selection bias related to streaks.\footnote{In this panel regression framework, the bias from introducing fixed-effects is an example of an incidental parameter problem of \citet{NeymanScott--Econometrica--1948}, and is essentially equivalent to that discussed in \citet{Nerlove--Econometrica--1971} and \citet{Nickell--Econometrica--1981}, and itself is closely related to the bias in estimates of autocorrelation in time series mentioned in the Introduction.} As a result, a bias correction is necessary.  In this case, the estimated effect is +13.9 percentage points ($p<.01$, $S.E.=5.8$), which has larger standard errors because the heteroscedasticity under the assumption of different player probabilities necessitates the use of robust variants (in this case, Bell and McCaffrey standard errors,\setcitestyle{square}see  \citet{ImbensKolesar--WP--2016}).\setcitestyle{round}The magnitude of the estimated effect has a different interpretation than the one given for the estimate of the average difference across players; it should be thought of as the hot hand effect for the average shot rather than the average player.  This interpretation arises because pooling shots across players generates an unbalanced panel, which results in the  estimate placing greater weight on the players that have taken more shots. As such, in the extreme it is even possible that the majority of players exhibit a tendency to have fewer streaks than expected by chance, yet, because they have generated relatively few observations, their data becomes diluted by many observations from a single streak shooter.

\subsection{Details on the hypothesis testing with the permutation test procedure\label{ssec: Appendix Permutation Details}}
Let $\bm{X}\in \{0,1\}^n$ be a sequence of shot outcomes from some player, $i$.  The null hypothesis is that the shots are i.i.d. with $\prob(X_t=1)=p^i$. This implies that conditional on the number of hits, $N_1(\bm{X})=n_1$, each rearrangement is equally likely. Considering only sequences  for which both $\hat{P}^i(hit| k\ hits)$ and  $\hat{P}^i(hit| k\ misses)$  are defined, the hot hand hypothesis predicts that the difference $\hat{P}^i(hit| k\ hits)-\hat{P}^i(hit| k\ misses)$ will be significantly larger than what one would expect by chance. Let $\hat{D}_k(\bm{X})$ be this difference for sequence $\bm{X}$. For an observed sequence $\bm{x}$, with $N_1(\bm{x})=n_1$ hits, to test the null hypothesis at the $\alpha$ level, one simply checks if $\hat{D}_k(\bm{x})\geq c_{\alpha,n_1}$, where the critical value $c_{\alpha,n_1}$ is defined as the smallest $c$ such that $\prob(D_k(\bm{X})\geq c \ |\ H_0,\ N_1(\bm{X})= n_1 ) \leq\alpha$, and the distribution $\prob(D_k(\bm{X})\geq c \ |\ H_0,\ N_1(\bm{X}) = n_1 )$ is generated using the formula provided in Web Appendix~\ref{ssec: Appendix recursive diff}. For the quantity $\prob(D_k(\bm{X})\geq c \ |\ H_0,\ N_1(\bm{X})= n_1 )$ it may be the case that for some $c^*$, it is strictly greater than $\alpha$ for $c\leq c^*$, and equal to $0$ for $c>c^*$. In this case, for any sequence with $N_1(\bm{X})= n_1 $ one cannot reject $H_0$ at an $\alpha$ level of significance. From the ex ante perspective, a test of the hot hand at the $\alpha$ level of significance consists of a family of such critical values $\{c_{\alpha,n_1}\}$.  It follows immediately that $\prob(\text{reject}|H_0)\leq \alpha$ because $\prob(\text{reject}|H_0)=\sum_{n_1=1}^{n} \prob(D_k(\bm{X})\geq c_{\alpha,n_1} | H_0,\ N_1(\bm{X}) = n_1)\prob( N_1(\bm{X}) = n_1|H_0) \leq \alpha$.  Lastly, for any arbitrary test statistic $T(\bm{X})$, the fact that the distribution of $\bm{X}$ is \emph{exchangeable} conditional on $N_1(\bm{X}) = n_1$ means that $\prob(T(\bm{X})\geq c \ |\ H_0,\ N_1(\bm{X})= n_1 )$ can be approximated to appropriate precision with Monte-Carlo permutations of the sequence $\bm{x}$.

\clearpage
\section{Web Appendix: Streak Selection Bias and a Quantitative Comparison to Sampling without Replacement}\label{sec: Web Appendix bias Bias Mechanism}

We show how the downward bias in the estimator $\hat{P}_k(X)$ is driven by two sources of selection bias. One is related to sampling-without-replacement, and the other to the overlapping nature of streaks.

Recall from the proof of Theorem~\ref{thm: General bias} that $E\left.\left[ \hat{P}_{k}(\bm{X})\right| I_{k}(\bm{X}) \neq \emptyset \right]=\prob(X_\tau=1| I_{k}(\bm{X}) \neq \emptyset)$, where $\tau$ is drawn (uniformly) at random from $I_k(\bm{X})$.  Because any sequence $\bm{X}\in\{0,1\}^n$, such that $I_{k}(\bm{X})\neq \emptyset$, that a researcher encounters will contain a certain number of successes $N_1(\bm{X})=n_1$ and failures $n_0:=n-n_1$, for $n_1=k,\dots,n$ we can write $\prob(X_\tau=1| I_{k}(\bm{X}) \neq \emptyset)=\sum_{n_1=k}^n \prob(X_\tau=1|  N_1(\bm{X})=n_1, I_{k}(\bm{X}) \neq \emptyset)\prob(N_1(\bm{X})=n_1|I_{k}(\bm{X}) \neq \emptyset)$.  To explore the nature of the downward bias we discuss why $\prob(X_\tau=1|  N_1(\bm{X})=n_1, I_{k}(\bm{X}))<\prob(X_t=1|  N_1(\bm{X})=n_1)=n_1/n$, i.e. why the probability that a randomly drawn trial from $I_{k}(\bm{X})$ is less than the overall proportion of successes in the sequence $\hat{p}=n_1/n$, i.e. the prior probability that a trial is a success when it is drawn (uniformly) at random from $1,\dots,n$ under the knowledge that $N_1(\bm{X})=n_1$.\footnote{Note that $\prob(N_1(\bm{X})=n_1|I_{k}(\bm{X}) \neq \emptyset)>\prob(N_1(\bm{X})=n_1)$ because the exclusion of sequences without a streak of $k$ successes in the first $n-1$ trials biases upwards the number of successes.  We do not consider this upward bias here as Theorem~\ref{thm: General bias} shows that the downward biases predominate.}

Suppose that the researcher were to know the overall proportion of successes $\hat{p}=n_1/n$ in the sequence. Now, consider the following two ways of learning that trial $t$ immediately follows $k$ consecutive successes: (i) a trial $\tau_N$, drawn uniformly at random from $\{k+1,\dots, n\}$  is revealed to be trial $\tau_N=t$, and preceded by $k$ consecutive successes, or (ii) a trial $\tau_I$, drawn (uniformly) at random from $I_{k}(\bm{X})=\{i:\prod_{i=t-k}^{t-1}X_i=1\}\subseteq \{k+1,\dots, n\}$ is revealed to be trial $\tau_I=t$. In each case, the prior probability of success is $\prob(X_t=1)=n_1/n$, which can be equivalently represented with the odds ratio $\prob(X_t=1)/\prob(X_t=0)=n_1/n_0$, indicates the $n_1/n_0:1$ prior odds in favor of $X_t=1$ (relative to $X_t=0$).

In the first case the probability distribution for $\tau_N$ is given by  $\prob(\tau_N=t)=1/(n-k)$ for all $t\in \{k+1,\dots, n\}$, and is independent of $\bm{X}$. Upon finding out that $\tau_N=t$ one then learns that $\prod_{t-k}^{t-1}X_i=1$. As a result, the posterior odds can be represented by a sampling-without-replacement formula, via Bayes rule:

\begin{align*}
\frac{\prob(X_t=1|\tau_N =t)}{\prob(X_t=0|\tau_N =t)}&=\frac{\prob(X_t=1,\prod_{t-k}^{t-1}X_i=1|\tau_N =t)}{\prob(X_t=0,\prod_{t-k}^{t-1}X_i=1|\tau_N =t)}\\
&=\frac{\prob(\tau_N=t|X_t=1,\prod_{t-k}^{t-1}X_i=1)}{\prob(\tau_N=t|X_t=0,\prod_{t-k}^{t-1}X_i=1)}\frac{\prob(\prod_{t-k}^{t-1}X_i=1| X_t=1)}{\prob(\prod_{t-k}^{t-1}X_i=1| X_t=0)}\frac{\prob(X_t=1)}{\prob(X_t=0)}\\
&=\frac{\prob(\prod_{t-k}^{t-1}X_i=1| X_t=1)}{\prob(\prod_{t-k}^{t-1}X_i=1| X_t=0)}\frac{\prob(X_t=1)}{\prob(X_t=0)}\\
&=\frac{\frac{n_1-1}{n-1}\times\cdots\times\frac{n_1-k}{n-k}}{\frac{n_1}{n-1}\times\cdots\times\frac{n_1-k+1}{n-k}}\frac{n_1}{n_0}\\
&=\frac{n_1-k}{n_1}\frac{n_1}{n_0}\\
&=\frac{n_1-k}{n_0}
\end{align*}
Observe that the prior odds in favor of success are attenuated by the likelihood ratio $\frac{n_1-k}{n_1}$ of producing $k$ consecutive successes given either hypothetical state of the world: $X_t=1$ or $X_t=0$, respectively.  That this is a sampling-without-replacement effect can be made most transparent by re-expressing the posterior odds as $\frac{n_1-k}{n-k}/\frac{n_0}{n-k}$.\footnote{The numerator is the probability of drawing a 1 at random from an urn containing $n_1$ 1's and $n_0$ 0's, once $k$ 1's (and no 0's) have been removed from the urn. The denominator is the probability of drawing a 0 from the same urn.}\fnsep\footnote{This effect calls to mind the key behavioral assumption made in \cite{Rabin-QJE-2002}, that believers in the law of small numbers view outcomes from an i.i.d. process as if they were instead generated by random draws without replacement.}

In the second case, the probability that $\tau_I=t$ is drawn from $I_{k}(\bm{X})$ is completely determined by $M:=|I_{k}(\bm{X})|$, and equal to $1/M$. Upon learning that $\tau_I=t$ one can infer the following three things: (i) $I_{k}(\bm{X})\neq \emptyset$, i.e. $M\geq 1$, which is informative if $n_1\leq (k-1)(n-n_1)+k$, (ii) $t$ is a member of $ I_{k}(\bm{X})$, and (iii) $\prod_{t-k}^{t-1}X_i=1$, as in sampling-without-replacement. As a result, the posterior odds can be determined via Bayes Rule in the following way:
\begin{align}
\frac{\prob(X_t=1|\tau_I =t)}{\prob(X_t=0|\tau_I =t)}&=\frac{\prob(X_t=1,\prod_{t-k}^{t-1}X_i=1,M \geq 1|\tau_I =t)}{\prob(X_t=0,\prod_{t-k}^{t-1}X_i=1,M \geq 1|\tau_I =t)}\nonumber\\
&=\frac{\prob(\tau_I=t|X_t=1,\prod_{t-k}^{t-1}X_i=1,M\geq 1)}{\prob(\tau_I=t|X_t=0,\prod_{t-k}^{t-1}X_i=1,M\geq 1)}\frac{\prob(X_t=1,\prod_{t-k}^{t-1}X_i=1,M\geq 1)}{\prob(X_t=0,\prod_{t-k}^{t-1}X_i=1,M\geq 1)}\nonumber\\
&=\frac{\prob(\tau_I=t|X_t=1,\prod_{t-k}^{t-1}X_i=1)}{\prob(\tau_I=t|X_t=0,\prod_{t-k}^{t-1}X_i=1)}\frac{\prob(\prod_{t-k}^{t-1}X_i=1| X_t=1)}{\prob(\prod_{t-k}^{t-1}X_i=1| X_t=0)}\frac{\prob(X_t=1)}{\prob(X_t=0)}\label{eq: SWOR1 appendix}\\
&=\frac{E\left.\left[\ \frac{1}{M}\ \right|\ \prod_{t-k}^{t-1}X_i=1, X_t=1\right] }{E\left.\left[\ \frac{1}{M}\ \right|\ \prod_{t-k}^{t-1}X_i=1, X_t=0\right] }\frac{\prob(\prod_{t-k}^{t-1}X_i=1| X_t=1)}{\prob(\prod_{t-k}^{t-1}X_i=1| X_t=0)}\frac{\prob(X_t=1)}{\prob(X_t=0)}\nonumber\\
&=\frac{E\left.\left[\ \frac{1}{M}\ \right|\ \prod_{t-k}^{t-1}X_i=1, X_t=1\right] }{E\left.\left[\ \frac{1}{M}\ \right|\ \prod_{t-k}^{t-1}X_i=1, X_t=0\right] }\frac{n_1-k}{n_1}\frac{n_1}{n_0}\label{eq: SWOR2 appendix}
\end{align}
For first term in (\ref{eq: SWOR1 appendix}) the event $M\geq 1$ is dropped from the conditional argument because it is implied by the event $\prod_{t-k}^{t-1}X_i=1$, and the term $\frac{\prob(M \geq 1|X_t=1, \prod_{t-k}^{t-1}X_i=1)}{\prob(M \geq 1|X_t=0, \prod_{t-k}^{t-1}X_i=1)}$ does not appear because it is equal to 1.

Equation~\ref{eq: SWOR2 appendix} gives the posterior odds $\frac{\prob(X_t=1|\tau_I =t)}{\prob(X_t=0|\tau_I =t)}$ in favor of observing $X_t=1$ (relative to $X_t=0$), for a representative trial $\tau=t$ drawn at random from $I_{k}(\bm{X})$. Observe that the prior odds ratio $n_1/n_0$ is multiplied by two separate updating factors, which we now discuss.

The first updating factor $\frac{n_1-k}{n_1}$ is clearly strictly less than one and reflects the restriction that the finite number of available successes  places on the procedure for selecting trials into $I_{k}(\bm{X})$. In particular, it can be thought of as the information provided upon learning that $k$  of the $n_1$ successes are no longer available, which leads to a sampling-without-replacement effect on the prior odds $n_1/n_0$. Clearly, the attenuation in the odds due to this factor increases in the streak length $k$.

The second updating factor $\frac{E\left.\left[\ \frac{1}{M}\ \right|\ \prod_{t-k}^{t-1}X_i=1, X_t=1\right] }{E\left.\left[\ \frac{1}{M}\ \right|\ \prod_{t-k}^{t-1}X_i=1, X_t=0\right] }<1$, for $t<n$, reflects an additional restriction that the arrangement of successes and failures in the sequence places on the procedure for selecting trials into $I_{k}(\bm{X})$. It can be thought of as the additional information provided by learning that the $k$ successes, which are no longer available, are consecutive and immediately precede $t$.  To see why the odds are further attenuated in this case, we begin with the random variable $M$, which is defined as the number of trials in $I_{k}(\bm{X})$. The probability of any particular trial $t \in I_{k}(\bm{X})$ being selected at random is $1/M$.  Now, because the expectation in the numerator conditions on $X_t=1$, this means that $1/M$ is expected to be smaller in the numerator than in the denominator, where the expectation instead conditions on $X_t=0$. The reason why is that for a sequence in which $X_t=1$ the streak of 1's continues on, meaning that trial $t+1$ must also be in $I_{k}(\bm{X})$, and trials $t+2$ through $t+k$ each may also be in $I_{k}(\bm{X})$. By contrast, for a sequence in which $X_t=0$ the streak of 1's ends, meaning that trials $t+1$ through $t+k$ cannot possibly be in $I_{k}(\bm{X})$, which leads the corresponding $1/M$ to be smaller in expectation.\footnote{This is under the assumption that $t\leq n-k$. In general, the event $X_t=0$ excludes the next $\min\{k,n-t\}$ trials from $t+1$ to $\min\{t+k,n\}$ from being selected, while the event $X_t=1$ leads trial $t+1$ to be selected, and does not exclude the next $\min\{k,n-t\}-1$ trials from being selected.} This last argument provides intuition for why the attenuation of the odds due to this factor increases in $k$.

\begin{figure}[t]
\begin{centering}
  \includegraphics[height=4.5in]{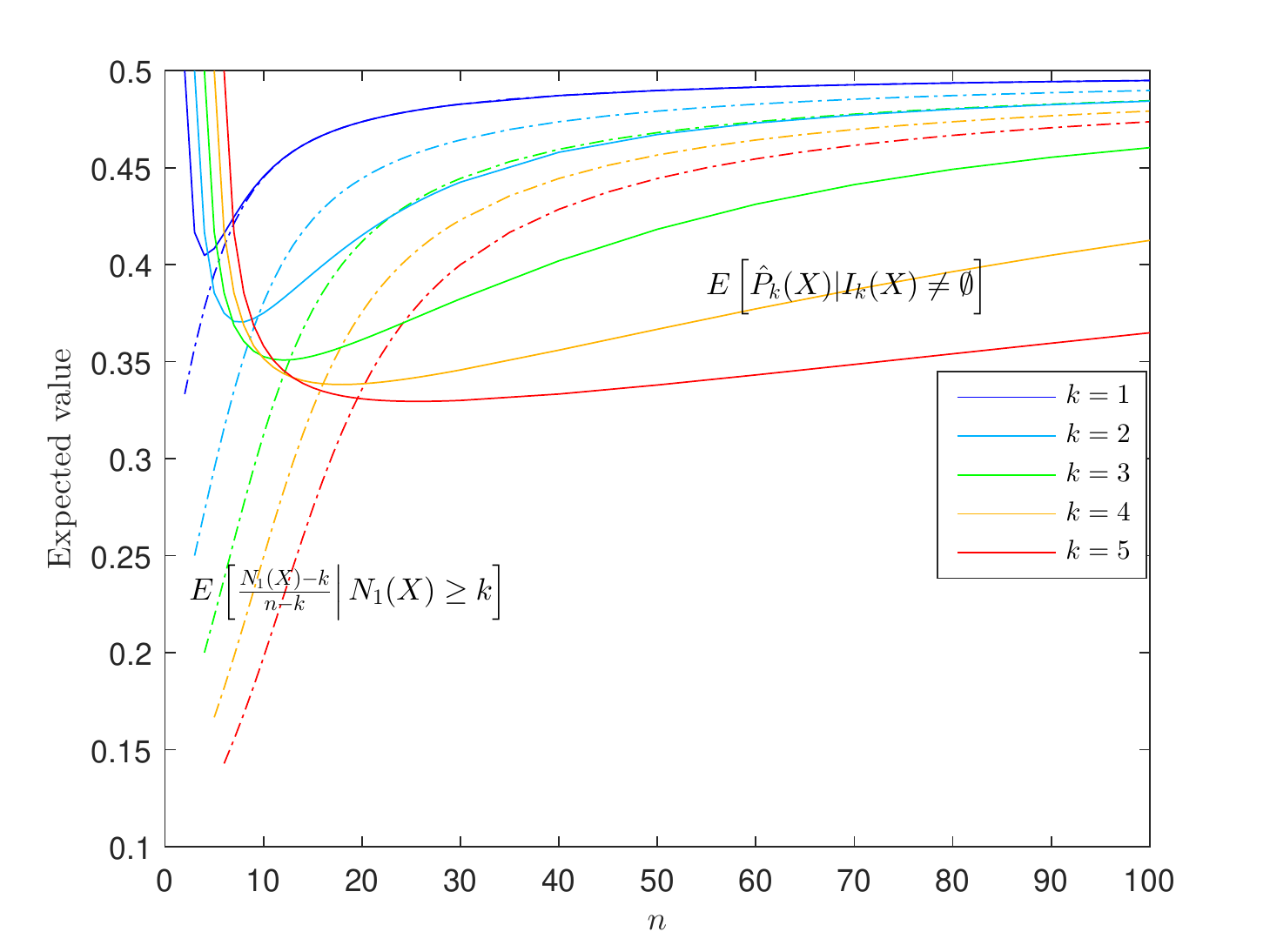}
  \caption{The dotted lines correspond to the bias from sampling-without-replacement. It is the expected probability of a success, given that $k$ successes are first removed from the sequence (assuming $p=.5$). The solid lines correspond to the expected proportion from Figure~\ref{fig: Bias Prop}.  }
  \label{fig: SWOR}
\end{centering}
\end{figure}

Interestingly, for the special case of $k=1$, $\frac{E\left.\left[\ \frac{1}{M}\ \right|\ x_{t-1}=1, x_t=1\right] }{E\left.\left[\ \frac{1}{M}\ \right|\ x_{t-1}=1, x_t=0\right] }=1-\frac{1}{(n-1)(n_1-1)}<1$ when $t<n$, and $\frac{E\left.\left[\ \frac{1}{M}\ \right|\ x_{n-1}=1, x_n=1\right] }{E\left.\left[\ \frac{1}{M}\ \right|\ x_{n-1}=1, x_n=0\right] }=\frac{n_1}{n_1-1}>1$ when $t=n$.\footnote{The likelihood ratios can be derived following the proof of Lemma~\ref{lemma: E(P11|N1)} in Appendix~\ref{sec: Appendix k=1 Proofs}. In particular, for the equivalent likelihood ratio, $\frac{\prob(\tau =t|x_{t-1}=1,x_t=1)}{\prob(\tau =t|x_{t-1}=1,x_t=0)}$, the approach used to derive the numerator can also be used to show that the denominator is equal to $\frac{1}{n-2}\left(\frac{n_0-1}{n_1} + \frac{n_1-1}{n_1-1}\right)$. Further, in the case of $t=n$, it is clear that $\prob(\tau =n|x_{n-1}=1,x_n=0)=\frac{1}{n_1}$. Each likelihood ratio then follows from dividing and collecting terms.}
These contrasting effects combine to yield the familiar sampling-without-replacement formula:
\begin{equation}
E\left.\left[\ \hat{P}_{1}(\bm{X})\ \right| \ I_{1}(\bm{X}) \neq \emptyset  ,\ N_1(\bm{X})=n_1 \ \right] = \frac{n_1-1}{n-1}\label{eq: SWOR prop k=1}
\end{equation}
as demonstrated in Lemma~\ref{lemma: E(P11|N1)}, in Appendix~\ref{sec: Appendix k=1 Proofs}. On the other hand, when $k>1$ the bias is substantially stronger than sampling-without-replacement (see Figure~\ref{fig: SWOR}), though the formula does not appear to admit a simple representation. For further discussion on the relationship between the bias, sampling-without-replacement, and the {\it overlapping words paradox} \citep{GuibasOdlyzko--JCTA--1981} see Web Appendix~\ref{sec: Appendix relationship with known biases}.

\paragraph{A quantitative comparison with sampling-without-replacement}

For the general case, in which  $\hat{p}=n_1/n$ is unknown,  juxtaposing the bias with sampling-without-replacement puts the magnitude of the bias into context. Let the probability of success be given by $p=\prob(X_t=1)$. In Figure~\ref{fig: SWOR}, the expected empirical probability that a randomly drawn trial in $I_{k}(\bm{X})$ is a success, which is the expected proportion, $E[\hat{P}_{k}(\bm{X})\ | \ I_{k}(\bm{X}) \neq \emptyset ]$, is plotted along with the expected value of the probability that a randomly drawn trial $t\in \{1,\dots,n\}\backslash T_k$ is a success, given that the $k$ success trials $T_k\subseteq \{1,\dots,n\}$ have already been drawn from the sequence (sampling-without-replacement), $E\left.\left[\frac{N_1(\bm{X})-k}{n-k} \right| N_1(\bm{X})\geq k\right]$. The plot is generated using the combinatorial results discussed in Section~\ref{ssec: Quantifying the bias}. Note that in the case of $k=1$, the bias is identical to sampling-without-replacement, as shown in Equation~\ref{eq: SWOR prop k=1}.\footnote{This appears to contradict equation~\ref{eq: SWOR2 appendix}, i.e. that the bias in the procedure used to select the subset of trials $I_{1k}(\bm{X})$, is {\it stronger} than sampling-without-replacement for $t<n$, whereas it is non-existent (thus weaker) for $t=n$.  This disparity is due to the second updating factor, which relates to the arrangement. It turns out that for $k=1$, the determining aspect of the arrangement that influences this updating factor is whether or not the final trial is a success, as this determines the number of successes in the first $n-1$ trials, where $M=n_1-X_n$. If one were to instead fix $M$ rather than $n_1$, then sampling-without-replacement relative to the number of successes in the first $n-1$ trials would be an accurate description of the mechanism behind the bias, and it induces a negative dependence between any two trials within the first $n-1$ trials of the sequence. Therefore, it is sampling-without-replacement with respect to $M$ that determines the bias when $k=1$.} Observe that for $k>1$, and $n$ not too small, the bias in the expected proportion is considerably larger than the corresponding bias from sampling-without-replacement.

\clearpage
\newpage
\section{Web Appendix: The Formula Used to Generate the Sampling Distribution and Calculate Expectations}\label{sec: Appendix recursive}

We describe the formula used to build the exact sampling distribution of the proportion, and difference in proportions, from which we calculate expectations and plot histograms.

\subsection{Proportion}\label{ssec: Appendix recursive proportion}

Given $n$ trials and streaks of length $k$, we observe that the proportion of successes on the trials that immediately follow $k$ consecutive success  $\hat{P}_{k}(\bm{x})$ can be represented simply as the the number of successes on trials that immediately follow a streak of $k$ consecutive successes divided by the total number of trials---i.e. failures and success---that immediately follow a streak of $k$ consecutive successes. In particular, for a sequence $\bm{x}\in\{0,1\}^n$ of successes and failures, we have:

\begin{align*}
\hat{P}_{k}(\bm{x})=\frac{M^1(\bm{x})}{M^0(\bm{x})+M^1(\bm{x})}
\end{align*}
where $M^0(\bm{x}):=|\{i\in\{k+1,\dots,n\}: (1-x_i)\prod_{j=i-k}^{i-1} x_{j}=1\}|$ is the number of failures that immediately follow $k$ consecutive successes (suppressing the $k$ to ease notation). Similarly, the number of successes that immediately follow $k$ consecutive successes is defined as $M^1(\bm{x}):=\{i\in\{k+1,\dots,n\}: x_i\prod_{j=i-k}^{i-1} x_{j}=1\}$.  Finally, the expected value of $\hat{P}_{k}(\bm{x})$ is uniquely determined by the joint distribution of counts  $\prob(\ (M^0(\bm{X}),M^1(\bm{X}))=(m^0,m^1)\ )$.

\begin{table}[h]
\begin{minipage}{1\textwidth}
\caption{In the table to the left column one lists the sample space of eight possible sequence realizations from three trials. Column two lists the number of (failures, successes) that immediately follow a success. The third column lists the probability with which the sequence occurs, where $p$ is the probability of success and $q$ is the probability of failure. In the table to the right the joint distribution is represented as a dictionary of count-probability pairs. Each unique count $\bm{m}=(m^0, m^1)$ has a unique associated probability equal to the sum of the probabilities of all sequences with the same associated count (see the table on the left).\label{tab: dictionary}}
\end{minipage}

\begin{minipage}{0.6\textwidth}
\centering\begin{tabular}{ccc}
\multicolumn{3}{c}{\emph{Sample Space of sequences}}\\
  \toprule
  sequence   & probability &  count\\
  \midrule
  $000$  & $q^3$  &  $(0,0)$\\
  $001$  & $q^2p$ &  $(0,0)$\\[3pt]
  $010$  & $q^2p$ &  $(1,0)$\\[3pt]
  $100$  & $q^2p$ &  $(1,0)$\\[3pt]
  $011$  & $qp^2$ &  $(0,1)$\\[3pt]
  $101$  & $qp^2$ &  $(1,0)$\\[3pt]
  $110$  & $qp^2$ &  $(1,1)$\\[3pt]
  $111$  & $p^3$  &  $(0,2)$\\
  \bottomrule
\end{tabular}
\end{minipage}
\begin{minipage}{0.4\textwidth}
\centering
\begin{tabular}{ccc}
\multicolumn{3}{c}{\emph{Dictionary}}\\
  \toprule
  count &   & probability\\
   $\bm{m}$ &  : &$p_D(\bm{m})$\\
  \midrule
  $(0,0)$ & : & $q^2$ \\
  $(1,0)$ &  : & $(q+q^2)p$ \\[3pt]
  $(0,1)$ &  : & $qp^2$ \\[3pt]
  $(1,1)$ &  : & $qp^2$ \\[3pt]
  $(0,2)$ &  : & $p^3$ \\
  \bottomrule
\end{tabular}
\end{minipage}
\end{table}

The algorithm described below (recursively) constructs the exact joint distribution of counts, by associating each unique count realization, which we call a \emph{key}, with its corresponding probability.\footnote{This algorithm, which builds upon an algorithm suggested by Michael J. Wiener, replaces an exact formula based on the joint distribution of runs of various lengths that we derived in a previous working paper version of this manuscript. The previous formula, while numerically tractable, was less efficient.}  In general, for a sequence of length $n$ and a streak of length $k$ this joint distribution can be represented as a \emph{dictionary} of (key:probability) pairs $D:=(\bm{m}: p_D(\bm{m}))_{\bm{m}\in D_c}$, where $\bm{m}:=(m^0, m^1)$ is a unique pair, $D_c$ corresponds to the set of count realizations with non-zero probability, i.e.  \[D_c:=\{\bm{m}\in\mathbb{N}^2 \ | \  p_D(\bm{m})>0\}
\]
and $p_D(\bm{m}):=\prob(\ (M^0(\bm{X}),M^1(\bm{X}))=(m^0,m^1)\ )$.

Table~\ref{tab: dictionary} reports the distribution over the sample space of sequences, and the corresponding dictionary, for the simple case of $n=3$ and $k=1$. From the dictionary one can derive the sampling distribution of the proportion and directly compute the expected proportion:
\[
E[\hat{P}_{k}(\bm{x}) | I_k(\bm{x})\neq\emptyset]= \sum_{\bm{m}\in D^*_c}\frac{m^1}{m^0+m^1}p^*_D(\bm{m})\]
where $D^*_c=D_c\setminus\{(0,0)\}$ and $p^*_D(\bm{m}):=p_D(\bm{m})/\sum_{\bm{m}' \in D^*_c} p_D(\bm{m}')$

Let $D(\ell,r)$ be the dictionary that represents the count-probability pairs for the remaining $r$ trials of a sequence that has $\ell\leq k$ consecutive successes immediately preceding the current trial. For example, if $k=2$ then $D(0,0)=D(1,0)=D(2,0)=((0,0):1)$, as when zero trials remain in the sequence the only count possible is (0,0), which occurs with probability 1.  Also note that $D(1,1)=((0,0):1)$, $D(2,1)=( (1,0):q, (0,1):p)$, and $D(2,2)=( (1,0):q, (1,1):pq, (0,2):p^2)$,
as a trial can only be counted as a fail or success if it is immediately preceded by $\ell=k=2$ consecutive successes. The key observation is that given the initial condition $D(\ell,0)=((0,0):1)$ for $0\leq\ell\leq k$, the dictionaries $D(\ell,r)$ can be defined recursively for $r>0$ and $0\leq\ell\leq k$, and take the following form:

\[
D(\ell,r)=
\begin{cases}
D(0,r-1)^{(0,0):q} \uplus D(\ell+1,r-1)^{(0,0):p}, & \quad \text{if } \ell<k  \\
D(0,r-1)^{(1,0):q} \uplus D(k,r-1)^{(0,1):p}, &\quad \text{if } \ell=k\\
\end{cases}
\]
where: (i) the operation $D^{\bm{m}':p'}:=(\bm{m}+\bm{m}':p_D(\bm{m})\times p')_{\bm{m}\in D_c}$ increments each count $\bm{m}$ with the addition of $\bm{m}'$, and scales its corresponding probability $p_D(\bm{m})$ by the probability $p'$ of the increment, and (ii) given the dictionaries $A$ and $B$, the operation $A \uplus B:=(\bm{m}:(p_A+p_B)(\bm{m}))_{\bm{m}\in A_c \cup B_c}$ defines the union of two dictionaries as the union of their counts, where the corresponding probabilities for a key that appears in both dictionaries are summed together (we assume that $p_A(\bm{m})=0$ for  $\bm{m}\notin A_c$; also for $B$). If a trial is immediately preceded by $\ell<k$ consecutive successes, then with probability $q$ ($p$) the next trial to its right will be immediately preceded by $0$ ($\ell+1$) consecutive successes; regardless of the outcome of the trial, $\bm{m}'=(0,0)$ additional failures and successes will be counted as immediately preceded by $k$ successes and $r-1$ trials will remain. If, on the other hand, a trial is immediately preceded by $\ell=k$ consecutive successes (and there is at least one trial remaining, i.e. $r>0$), then with probability $q$ ($p$) the next trial to its right will be immediately preceded by $0$ ($k$) consecutive successes and we will count $\bm{m}'=(1,0)$ ($(0,1)$) additional failures and successes; regardless of the outcome of the trial, $r-1$ trials will remain.

%If a trial is immediately preceded by $\ell<k$ consecutive successes, then with probability $q$ ($p$) the next trial to its right will be immediately preceded by $0$ ($\ell+1$) consecutive successes; regardless of the outcome of the trial, $\bm{m}'=(0,0)$ additional failures and successes will be counted as immediately preceded by $k$ successes and $r-1$ trials will remain. If, on the other hand, a trial is immediately preceded by $\ell=k$ consecutive successes (and followed by at least one trial, i.e. $r>0$), then with probability $q$ ($p$) the next trial to its right will be immediately preceded by $0$ ($k$) consecutive successes and we will count $\bm{m}'=(1,0)$ ($(0,1)$) additional failures and successes; regardless of the outcome of the trial, $r-1$ trials will remain.

The algorithm that follows  describes the complete recursive procedure.

\begin{algorithm}\label{algorithm: dictionary}
\caption{Recursive formula that builds the collection of dictionaries $D$. Of interest are the dictionaries $D(0,n)$ for $n=k+1,\dots, N$ which correspond to the joint distribution of the total number of (successes, failures) that immediately follow $k$ consecutive successes in $n$ trials.}
\SetAlgoLined
\DontPrintSemicolon
\SetKwFunction{FMain}{Count\_Distribution}
\SetKwProg{Fn}{Function}{:}{}
 \Fn{\FMain{$N$, $k$, $p$}}{
       \tcc{For the definition of $D(\ell,r)$, $A^{\bm{m}':p'}$ and $A \uplus B$ below, see text.}
 $q \gets 1-p$\;
\For{$n=0,\dots,N$}{
    $L\gets \min\{k,n\}$\;
    \For{$\ell=L,\dots,0$}{
        $r\gets n-\ell$\;
        \uIf{$r = 0 $}{
            $D(\ell,r) \gets ( (0,0):1)$\;
        }
        \uElseIf{$r>0$}{
            \uIf{$\ell<k$}{
            $D(\ell,r)\gets D(0,r-1)^{(0,0):q} \uplus D(\ell+1,r-1)^{(0,0):p}$\;
            }
            \uElseIf{$\ell=k$}{
            $D(\ell,r)\gets D(0,r-1)^{(1,0):q} \uplus D(k,r-1)^{(0,1):p}$\;
            }
        }
    }
}
\KwRet D\;
}
\end{algorithm}

\newpage

\subsection{Difference in Proportions}\label{ssec: Appendix recursive diff}

The difference in proportions can be computed from a dictionary $D:=(\bm{m}: p_D(\bm{m}))_{\bm{m}\in D_c}$, where $D_c$ corresponds to the set of count realizations with non-zero probability i.e.
\[D_c:=\{\bm{m}\in\mathbb{N}^4 \ | \  p_D(\bm{m})>0\}
\]
and $p_D(\bm{m}):=\prob(\ (M^0_0(\bm{X}),M^1_0(\bm{X}),M^0_1(\bm{X}),M^1_1(\bm{X}))=(m^0_0,m^1_0,m^0_1,m^1_1)\ )$. The variables $M^0_1(\bm{X})$ and $M^1_1(\bm{X})$ yield the total number of failures and successes (respectively) on those trials that immediately follow a streak of $k$ successes, whereas $M^0_0(\bm{X})$ and $M^1_0(\bm{X})$ yield the total number of failures and successes (respectively) on those trials that immediately follow a streak of $k$ failures.

Let $D(\ell_0,\ell_1,r)$ be the dictionary that represents the count-probability pairs for the remaining $r$ trials of a sequence in which there are $\ell_0\leq k$ consecutive failures and $\ell_1\leq k$ consecutive successes on the immediately preceding trials (so that $\ell_0\ell_1=0$).  These dictionaries can be constructed recursively in a way similar to that shown in Web Appendix \ref{ssec: Appendix recursive proportion}:
\[
D(\ell_0,\ell_1,r)=
\begin{cases}
D(\ell_0+1,0,r-1)^{(0,0,0,0):q} \uplus D(0,\ell_1+1,r-1)^{(0,0,0,0):p}, & \quad \text{if } \max\{\ell_0,\ell_1\}<k  \\
D(k,0,r-1)^{(1,0,0,0):q} \uplus D(0,1,r-1)^{(0,1,0,0):p}, &\quad \text{if } \ell_0=k\\
D(1,0,r-1)^{(0,0,1,0):q} \uplus D(0,k,r-1)^{(0,0,0,1):p},&\quad \text{if } \ell_1=k \\
\end{cases}
\]

See supplementary materials for the corresponding code.

\clearpage

\section{Web Appendix: The relationship between the streak selection bias and known biases and paradoxes}\label{sec: Appendix relationship with known biases}

\subsection{Sampling-without-replacement and the bias for streaks of length $k=1$.}\label{ssec: Appendix SWOR, Berkson, FSB}

A brief inspection of Table~\ref{tab: bias} in Section~\ref{sec: Introduction} reveals how the dependence between the first $n-1$ flips in the sequence arises. In particular, when the coin is flipped three times, the number of Hs in the first $2$ flips determines the number of observations of flips that immediately follow an H.  Because TT must be excluded, the first two flips will consist of one of three equally likely sequences: HT, TH or HH. For the two sequences with a single H---HT and TH---if a researcher were to find an H within the first two flips of the sequence and then select the adjacent flip for inspection, the probability of heads on the adjacent flip would be 0, which is strictly less than the overall proportion of heads in the sequence. This can be thought of as a sampling-without-replacement effect. More generally, across the three sequences, HT, TH, and HH, the expected probability of the adjacent flip being a heads is $(0+0+1)/3=1/3$. This probability reveals the (negative) sequential dependence that exists between the first two flips of the sequence. Further, the same negative dependence holds for \emph{any two flips} in the first $n-1$ flips of a sequence of length $n$, {\it regardless of their positions}. Thus, when $k=1$ it is neither time's arrow nor the arrangement of flips within the sequence that determines the bias.

This same sampling-without-replacement feature also underlies a classic form of selection bias known as Berkson's bias (aka Berkson's paradox). \citet{Berkson--BB--1946} presented a hypothetical study of the relationship between two diseases that, while not associated in the general population, become negatively associated in the population of hospitalized patients. The cause of the bias is subtle: patients are hospitalized only if they have {\it at least one} of the two particular diseases. To illustrate, assume that someone from the general population has a given disease (Y=``Yes'') or does not (N=``No''), with equal chances. Just as in the coin flip example, anyone with neither disease (NN) is excluded, while a patient within the hospital population must have one of the three equally likely profiles: YN, NY, or YY. Thus, just as with the coin flips, the probability of a patient having another disease, given that he already has one disease, is $1/3$.

The same sampling-without replacement feature again arises in several classic conditional probability paradoxes. For example, in the Monty Hall problem the game show host inspects two doors, which can together be represented as one of three equally likely sequences GC, CG, or GG (G=``Goat,'' C=``Car''), then opens one of the G doors from the realized sequence. Thus, the host effectively samples G without replacement \citep{Selvin--AS--1975,Nalebuff--JEP1--1987,VosSavant--ParadeMagazine--1990}.\footnote{The same structure also appears in what is known as the boy-or-girl paradox \citep{MillerSanjurjo2015c}. A slight modification of the Monty-Hall problem makes it identical to the coin flip bias presented in Table~\ref{tab: bias}\setcitestyle{square}(see \citet{MillerSanjurjo2015c}).\setcitestyle{round}}

Sampling-without-replacement also underlies a well-known finite sample bias that arises in standard estimates of autocorrelation in time series data \citep{Yule--JRSS--1926,ShamanStine--JASA--1988}. This interpretation of finite sample bias, which does not appear to have been previously noted, allows one to see how this bias is closely related to those above. To illustrate, let $\bm{x}$ be a randomly generated sequence consisting of $n$ trials, each of which is an i.i.d. draw from some continuous distribution with finite mean and variance. For a researcher to compute the autocorrelation she must first determine its sample mean $\bar{x}$ and variance $\hat{\sigma}^2(\bm{x})$, then calculate the autocorrelation $\hat{\rho}_{t,t+1}(\bm{x})=\hat{cov}_{t,t+1}(\bm{x})/\hat{\sigma}^2(\bm{x})$, where $\hat{cov}_{t,t+1}(\bm{x})$ is the autocovariance.\footnote{The autocovariance is given by  $\hat{cov}_{t,t+1}(\bm{x}):=\frac{1}{n-1}\sum_{i=1}^{n-1} (x_i-\bar{x})(x_{i+1}-\bar{x})$. } The total sum of values $n\bar{x}$ in a sequence serves as the analogue to the number of Hs (or Gs/Ys) in a sequence in the examples given above. Given $n\bar{x}$, the autocovariance can be represented as the expected outcome from a procedure in which one draws (at random) one of the $n$ trial outcomes $x_i$, and then takes the product of its difference from the mean $(x_i-\bar{x})$, and another trial outcome $j$'s difference from the mean. Because the outcome's value $x_i$ is essentially drawn from $n\bar{x}$, without replacement, the available sum total $(n\bar{x}-x_i)$ is averaged across the remaining $n-1$ outcomes, which implies that the expected value of another outcome $j$'s ($j\neq i$) difference from the mean is given by $E[x_j| x_i, \bar{x}]-\bar{x}=(n\bar{x}-x_i)/(n-1)-\bar{x}=(\bar{x}-x_{i})/(n-1)$. Therefore, given  $x_i-\bar{x}$, the expected value of the product $(x_i-\bar{x})(x_{j}-\bar{x})$  must equal $(x_i-\bar{x})(\bar{x}-x_{i})/(n-1)=-(x_i-\bar{x})^2/(n-1)$, which is independent of $j$. Because $x_i$ and $j$ were selected at random, this implies that the expected autocorrelation, given $\bar{x}$ and $\hat{\sigma}^2(\bm{x})$, is equal to $-1/(n-1)$ for all $\bar{x}$ and $\hat{\sigma}^2(\bm{x})$.  This result accords with known results on the $O(1/n)$ bias in discrete-time autoregressive processes \citep{Yule--JRSS--1926,ShamanStine--JASA--1988}, and happens to be identical to the result in Theorem~\ref{thm: E(D1)} for the expected difference in proportions (see Appendix~\ref{sec: Appendix k=1 Proofs}). In the context of time series regression this bias is known as the {\it Hurwicz bias} \citep{Hurwicz1950}, which is exacerbated when one introduces fixed effects into a time series model with a small number of time periods \citep{Nerlove1967,Nerlove--Econometrica--1971,Nickell--Econometrica--1981}.\footnote{The bias that is exacerbated by the introduction of of exogenous variables is commonly known as the ``Nickell bias,'' which was first explored by simulation by \citet{Nerlove1967,Nerlove--Econometrica--1971}. It is an example of what is known as the {\it incidental parameter problem} \citep{NeymanScott--Econometrica--1948,Lancaster--JE--2000}.}\fnsep\footnote{In a comment on this paper, \citet{RinottBar-Hillel--WP--2015} assert that the work of \citet{Bai--AnStat--1975} (and references therein) demonstrate that the bias in the proportion of successes on the trials that immediately follow a streak of $k$ or more successes follows directly from known results on the finite sample bias of Maximum Likelihood estimators of transition probabilities in Markov chains, as independent Bernoulli trials can be represented by a Markov chain with each state defined by the sequence of outcomes in the previous $k$ trials. While it is true that the MLE of the corresponding transition matrix is biased, and correct to note the relationship in this sense, the cited theorems do not indicate the direction of the bias, and in any event do not directly apply in the present case because they require that transition probabilities in different rows of the transition matrix not be functions of each other, and not be equal to zero, a requirement which does not hold in the corresponding transition matrix. Instead, an unbiased estimator of each transition probability will exist, and will be a function of the overall proportion.}\\

\subsection{Pattern overlap and the bias for streaks of length $k>1$.}\label{ssec: Web Appendix overlap k>1}

In Figure~\ref{fig: SWOR} of Web Appendix~\ref{sec: Web Appendix bias Bias Mechanism} we compare the magnitude of the bias in the (conditional) expected proportion to the pure sampling-without-replacement bias, in a sequence of length $n$. As can be seen, the magnitude of the bias in the expected proportion is nearly identical to that of sampling-without-replacement for $k=1$. However, for the bias in the expected proportion, the relatively stronger sampling-without-replacement effect that operates within the first $n-1$ terms of the sequence is balanced by the absence of bias for the final term.\footnote{The reason for this is provided in the alternative proof of Lemma~\ref{lemma: E(P11|N1)} in Appendix~\ref{sec: Appendix k=1 Proofs}}  On the other hand, for $k>1$ the bias in the expected proportion is considerably stronger than the pure sampling-without-replacement bias. One intuition for this is provided in the discussion of the updating factor in Section~\ref{sec: Web Appendix bias Bias Mechanism}. Here we discuss another intuition, which has to do with the overlapping nature of the selection criterion when $k>1$, which is related to what is known as the {\it overlapping words paradox} \citep{GuibasOdlyzko--JCTA--1981}.

For simplicity, assume that a sequence is generated by $n=5$ flips of a fair coin.  For the simple case in which streaks have length $k=1$, the number of flips that immediately follow a heads is equal to the number of instances of H in the first $n-1=4$ flips. For any given number of Hs in the first four flips, say three, if one were to sample an H from the sequence and then examine an adjacent flip (within the first four flips), then because any H could have been sampled, across all sequences with three Hs in the first four flips, any H appearing within the first four flips is given equal weight regardless of the sequence in which it appears. The exchangeability of outcomes across equally weighted sequences with an H in the sampled position (and three Hs overall) therefore implies that for any other flip in the first four flips of the sequence, the probability of an H is equal to $\frac{3-1}{4-1}=\frac{2}{3}$, regardless of whether or not it is an adjacent flip. On the other hand, for the case of streaks of length $k=2$, the number of opportunities to observe a flip that immediately follows two consecutive heads is equal to the number of  instances of HH in the first $4$ flips. Because the pattern HH can overlap with itself, whereas the pattern H cannot, then for a sequence with three Hs, if one were to sample an HH from the sequence and examine an adjacent flip within the first 4 flips, it is not the case that any two of the Hs from the sequence can be sampled. For example, in the sequence HHTH only the first two Hs can be sampled.  Because the sequences HHTH and HTHH each generate just one opportunity to sample, this implies that the single instance of HH within each of these sequences is weighted twice as much as any of the two (overlapping) instances of HH within the two sequences HHHT and THHH that each allow two opportunities to sample, despite the fact that each sequence has three heads in the first four flips. This implies that, unlike in the case of $k=1$, when sampling an instance of HH from a sequence with three heads in the first four flips, the remaining outcomes H and T are no longer exchangeable, as the arrangements HHTH and HTHH, in which every adjacent flip within the first four flips is a tails, must be given greater weight than the arrangements HHHT and THHH, in which half of the adjacent flips are heads.

This consequence of pattern overlap is closely related to the  {\it overlapping words paradox}, which states that for a sequence (string) of finite length $n$, the probability that a pattern (word) appears, e.g. \_HTTHH\_, depends not only on the length of the pattern relative to the length of the sequence, but also on how the pattern \emph{overlaps} with itself \citep{GuibasOdlyzko--JCTA--1981}.\footnote{For a simpler treatment which studies a manifestation of the paradox in the non-transitive game known as ``Penney's'' game, see \citet{Konold--AS--1995} and \citet{Nickerson--UMAP--2007}.}  For example, while the expected number of (potentially overlapping) occurrences of a particular two flip pattern---TT, HT,TH or HH---in a sequence of four flips of a fair coin does not depend on the pattern, it's probability of occurrence does.\footnote{That all fixed length patterns are equally likely ex-ante is straightforward to demonstrate. For a given pattern of heads and tails of length $\ell$, $(y_1,\dots,y_\ell)$, the expected number of occurrences of this pattern satisfies $E[\sum_{i=\ell}^n 1_{[(X_{i-\ell+1},\dots,X_i)=(y_1,\dots,y_\ell)]}]=\sum_{i=\ell}^n E[1_{[(X_{i-\ell+1},\dots,X_i)=(y_1,\dots,y_\ell)]}]=\sum_{i=\ell}^n 1/2^\ell= (n-\ell+1)/2^\ell$.} The pattern HH can overlap with itself, so can have up to three occurrences in a single sequence (HHHH), whereas the pattern HT cannot overlap with itself, so can have at most two occurrences (HTHT).  Because the expected number of occurrences of each pattern must be equal, this implies that the pattern HT is distributed across more sequences, meaning that any given sequence is more likely to contain this pattern.\footnote{Note that the proportion of heads on flips that immediately follow two consecutive heads can be written as the number of (overlapping) HHH instances in $n$ flips, divided by the number of (overlapping) HH instances in the first $n-1$ flips.}

\end{document}